\documentclass[aps,prl,reprint,superscriptaddress]{revtex4-2}

\usepackage{graphicx} 
\usepackage{dcolumn} 
\usepackage{bm} 
\usepackage[hidelinks]{hyperref}
\usepackage[mathlines]{lineno}

\usepackage{amsmath}
\usepackage{amssymb}
\usepackage{xcolor}

\begin{document}
	\title{Strong Correlation Drives Zero-Field Josephson Diode Effect}
	\author{Yiheng Sun}
	\thanks{These authors contributed equally to this work.}
	\affiliation{International Center for Quantum Design of Functional Materials (ICQD), University of Science and Technology of China, Hefei, Anhui 230026, China }
	\affiliation{Hefei National Laboratory, Hefei, Anhui 230088, China	}
	\author{Zhenyu Zhang}
	\thanks{These authors contributed equally to this work.}
	\affiliation{International Center for Quantum Design of Functional Materials (ICQD), University of Science and Technology of China, Hefei, Anhui 230026, China }
	\affiliation{Hefei National Laboratory, Hefei, Anhui 230088, China	}
	\author{James Jun He}
	\email{jun\_he@ustc.edu.cn}
	\affiliation{Hefei National Laboratory, Hefei, Anhui 230088, China	}
	\date{\today}
	
	\begin{abstract}

	The supercurrent diode effect (SDE), characterized by unequal critical currents in opposite directions, has been observed with or without magnetic fields, yet mechanisms enabling zero-field SDE without explicit symmetry breaking remain underexplored.
	Here we investigate a Josephson junction with strong electron-electron interaction modeled by a Hubbard \(U\) term and an odd number of electrons. We find that strong correlations induce spontaneous breaking of time-reversal and mirror symmetries, forming a \(\varphi\)-junction with degenerate energy minima at \(\pm\varphi\), resulting in zero-field Josephson diode effect (JDE) without magnetic order. Spin-orbit coupling breaks SU(2) symmetry but does not determine diode polarity, contrasting with magneto-chiral mechanisms. We further show that applying a tiny Zeeman field enables controllable JDE with sizable efficiency due to the enhancement by the strong magnetic correlation, and the JDE strength peaks when the field induces a level-crossing transition. These findings establish strong electron correlation as a distinct mechanism for nonreciprocal superconducting transport, broadening the understanding of SDE origins.
		
	\end{abstract}
	\maketitle

	\paragraph{Introduction ---} Nonreciprocal supercurrents have garnered significant interest recently, owing to their deep connections with fundamental quantum physics and their promising applications in energy-efficient electronics. This phenomenon, known as the supercurrent diode effect (SDE), is characterized by unequal critical currents $I_{c\pm}$ along opposite directions. 
	SDE has been observed both with \cite{Ando_Nature2020, hou2023PRL,gutfreund2023NC,li2024NC,baumgartner2022NN, pal2022NP,bauriedl2022NC,chen2024AFM,kim2024NC,guan2026CP} and without \cite{Wu_Nature2022,liu2024SA,ma2025CP, anwar2023CP,diez2023NC, trahms2023Nature,JShen2025arXiv, narita2022NN, lin2022NP, wan2024Nature,le2024nature,Qi_nature2025} the application of a magnetic field, in individual superconductors (SCs) \cite{Ando_Nature2020,narita2022NN,lin2022NP, hou2023PRL,wan2024Nature,le2024nature,Qi_nature2025,gutfreund2023NC,li2024NC} as well as in Josephson junctions (JJs) \cite{baumgartner2022NN, pal2022NP,bauriedl2022NC,anwar2023CP,chen2024AFM,guan2026CP, Wu_Nature2022,trahms2023Nature,diez2023NC,liu2024SA,kim2024NC,ma2025CP,JShen2025arXiv}. 
	
	The SDE has been observed across a wide range of material platforms \footnote{In the literature, {SDE} (superconducting diode effect) typically refers to the phenomenon in bulk superconductors while that in Josephson junctions is termed Josephson diode effect (JDE). Here, we use supercurrent diode effect (SDE) for both without distinction, given the profound similarity under interchange of Cooper pair momentum $q$ and Josephson phase $\phi$.}, stimulating diverse theoretical efforts to elucidate its underlying physics \cite{nadeem2023NRP,ma2025APR,zhang2022PRX}. Physical mechanisms for SDE in presence of external magnetic fields include superconducting magneto-chiral anisotropy (MCA) \cite{Yuan_pnas2022,Daido_PRL2022,He_NJP2022,tanaka2022PRB,lu2023PRL,he2023NC,Li_PRB2025}, asymmetric vortex motion \cite{Hou_PRL2023}, finite-momentum or inter-band pairing \cite{davydova2022SA,xie2023PRL,chakraborty2025prl, mei2025interband}, supercurrent interference \cite{ WZ_2012arXiv, CZ_PRB2018, souto_PRL2022},  etc.  
	Zero-field SDE can occur  in systems with intrinsic magnetism. If the magnetic order arises from local spins, it introduces an exchange field to the itinerant electrons, making the SDE mechanism similar to that under an external Zeeman field, where spin-orbit coupling (SOC) plays a key role. Alternatively, orbital magnetism can induce SDE independently of spin degrees of freedom, as demonstrated in Moire flat-band systems \cite{hu2023prl}, the Haldane model of quantum anomalous Hall insulators \cite{ShenPRB2025}, and potentially in kagome-lattice superconductors \cite{le2024nature}. 
	
	An exotic class of SDE happen without clear evidence of any magnetic order \cite{Wu_Nature2022,wan2024Nature,le2024nature,liu2024SA,ma2025CP,anwar2023CP,Qi_nature2025}. A possible origin of such a phenomenon is the formation of superconducting states that spontaneously break the time-reversal (TR) symmetry. Such a state may refer to an unconventional SC order parameter, such as chiral $p$-wave superconductivity \cite{zinkl2022prr}, or a JJ whose energy is lowest at a nonzero Josephson phase $\phi=\pm \varphi $. Such JJs are called $\varphi$-junctions \footnote{We follow the convention of Ref. \cite{buzdin2003prb}  to call a \text{Josephson} junction with double energy mimina at $\phi=\pm \varphi$  a \emph{$\varphi$-junction}, while that with a single minimum at $\phi=\varphi_0\neq  0 $ or  $ \pi $ a \emph{$\varphi_0$-junction}.}, of which one example is the $\pi/2$-junctions formed between twisted bilayers of $d$-wave SCs \cite{yang2018PRB,can2021NP,zhao2023Science,volkov2024PRB}.  

	Previous theoretical studies have often interpreted SDE within mean-field or phenomenological frameworks. While Coulomb interaction was considered in Refs. \cite{hu2007prl,misaki2021prb}, where nonreciprocity was found in the a.c. Josephson effect or in the re-trapping current $I_r$  (defined at the normal-superconducting transition), it remains an open question whether electron-electron interactions can lead to nonreciprocal d.c. critical current $I_c$, defined at the superconducting-normal transition.
	
	In this Letter, we investigate a Josephson junction where the Coulomb interaction in the normal region is modeled by Hubbard $U$  and treated exactly. We reveal a new mechanism for zero-field SDE in Josephson junctions (termed Josephson diode effect \cite{hu2007prl}, JDE) without magnetic order, in which strong electron correlation plays a crucial role and leads to a $\varphi$-junction when the electron parity is odd. The phase dynamics in such a $\varphi$-junction result in a d.c. JDE with unequal critical currents ($I_{c\pm}$). Notably, both TR symmetry ($\mathcal{T} $) and mirror symmetry $\mathcal{M}_x$ are broken spontaneously, in sharp contrast to existing theories of SDE where all symmetries that reverse the current direction must be explicitly broken.
	
\begin{figure}
		\includegraphics[width=1\columnwidth]{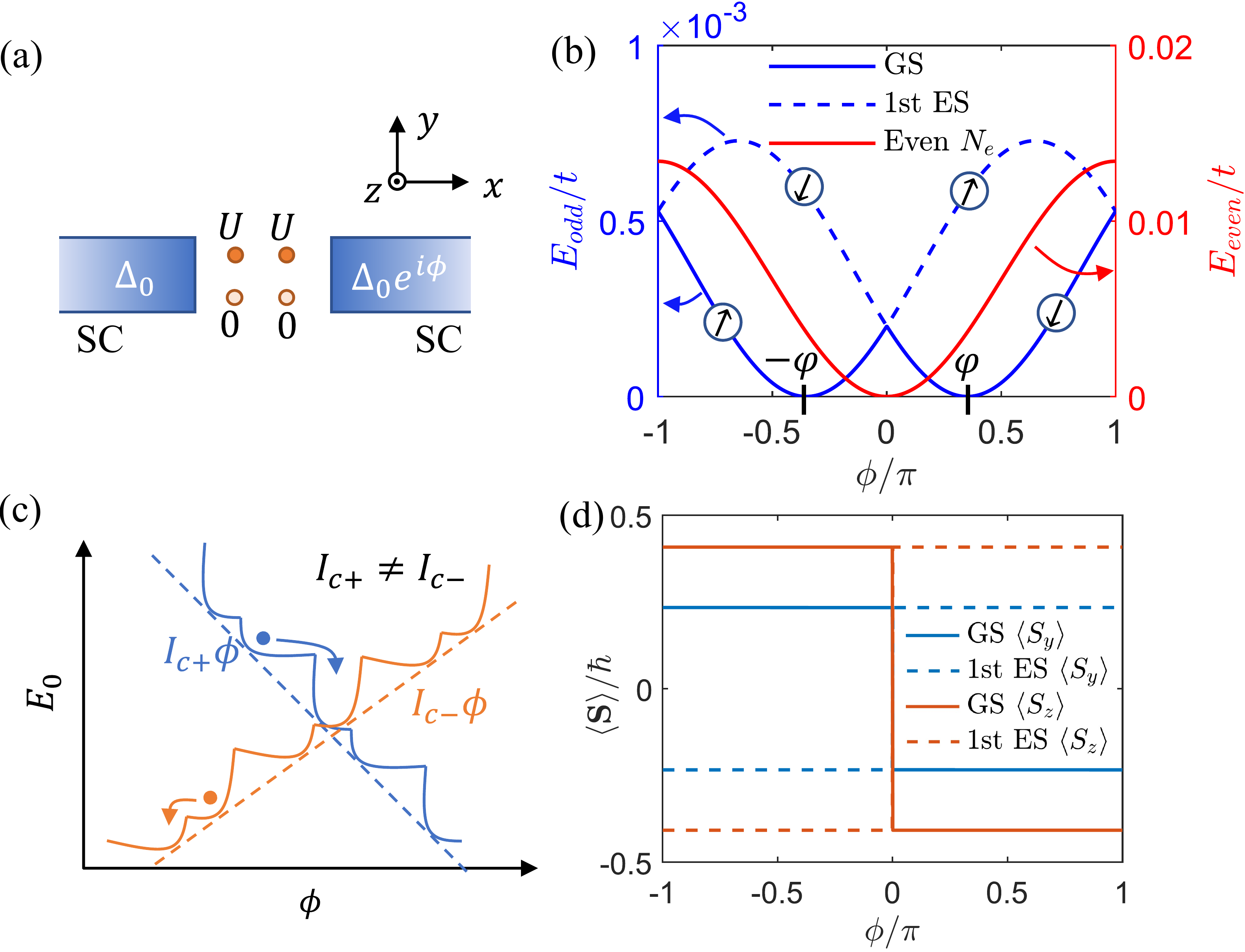} 
		\caption{(a) Schematics of a Josephson junction with in homogeneous onsite Coulomb repulsion. The Hubbard U on each site is labeled and a geometry with non-uniform U is chosen here. (b) The ground state (GS) energy as a function of the SC phase difference $\phi$ when the number of electrons $N_e$ is even ($\mu=0.8$, red curve, right axis) and when $N_e$ is odd ($\mu=1.2$, blue curves, left axis). (c) Illustration of the phase dynamics in a RCSJ model with a double-minimum potential. (d)  Spin polarization of the GS and the first excited state (1st ES) for $\mu=1.2$ as functions of the Josephson phase. Other parameters are $t=1, \mu=10, r=0.5, \Delta=0.1$, which apply throughout the manuscript unless stated otherwise.} 
		\label{fig:1}
	\end{figure}

	\paragraph{Model ---} 
    Consider the following Hamiltonian defined on a square lattice, 
	\begin{align}
		H = &\sum_{<ij>} [t \sum_{\sigma=\uparrow\downarrow} c^\dagger_{i\sigma} c_{j\sigma} + \Delta_i \delta_{ij} c^\dagger_{i\uparrow} c^\dagger_{j\downarrow} + h.c.  -\mu \delta_{ij} n_i ] \notag \\
		 &+ \sum_i U_i n_{i\uparrow }n_{i\downarrow} + H_{\text{SOC}}, \label{eq:1}
	\end{align}
	where $c_{i\sigma}$($c_{i\sigma}^\dagger$) annihilates (creates) an electron with spin $\sigma$ on site $i$. The onsite electron number operator $n_i=n_{i\uparrow}+n_{i\downarrow}$, and $n_{i\sigma}=c^\dagger_{i\sigma} c_{i\sigma}$. 
	The parameter $t$ denotes the hopping amplitude and $\mu$ is the chemical potential. 
	The subscript $ <ij>$ denotes the nearest neighbors and $h.c.$ stands for the Hermitian conjugate of previous terms. The second line of Eq. (\ref{eq:1}) contains the onsite Coulomb repulsion ($U_i \ge 0$) and the SOC.  The SOC form is not really important here as long as the SU(2) symmetry is broken. For simplicity, we take the Rashba SOC
	\begin{align}
		 H_{\text {SOC}} = &  \lambda \sum_{<ij>} \bm d_{ij} \times \bm \sigma_{\alpha \beta }c^\dagger_{i\alpha} c_{j\beta}.
		 \label{eq:2}
	\end{align}
	The $\bm d_{ij}$ connects site $j$ to site $i$ and $\bm \sigma $ is the Pauli-matrix-vector. 
	Hamiltonian (\ref{eq:1}) describes the normal region of a Josephson junction, as illustrated in FIG. \ref{fig:1} (a), where a minimal model with only four lattice sites is shown. The semi-infinite SC leads are integrated out and their effect is incorporated into the pairing terms proportional to $\Delta_i$, which has different phases, zero or $\phi $, depending on the SC in its proximity \cite{Affleck2000PRB,Tanaka2002PhysicaC,Bergeret2007PRB}. This treatment enables us to obtain the energy eigenstates  by exact diagonalization.

	The total number of electrons $N_e$ in the ground state (GS) can be tuned by $\mu$. When $N_e$ is even, the total spin is integral and thus the GS is generally not degenerate even though $\mathcal{T}$ is preserved. The first excited state (1st ES) is separated from the GS by a rather large energy spacing compared to the SC gap. When the SCs are proximitized, the GS is only weakly affected and the energy follows the conventional Josephson relation $E_0=-E_J \cos \phi $, as shown in FIG. \ref{fig:1} (b). 	

	\paragraph{$\varphi$-junction and zero-field JDE ---} 
	The situation with odd $N_e$ is drastically different. Since the total spin is a half integer, the GS must be degenerate at $\phi=0$ or $\pi$ where Eq. (\ref{eq:1}) respects TR symmetry. In this case, the energies of both the GS and  the 1st ES are shown as functions of $\phi$ in FIG. \ref{fig:1} (b). As $\phi$ deviates from $0$ or $\pi$, the TR symmetry is broken and the degeneracy is lifted. Remarkably, the lowest energy happens at some non-special phases $\phi_0=\pm \varphi$. The macroscopic GS of the Josephson junction will choose either of the two minima, spontaneous breaking the TR symmetry $\mathcal{T}$ and the mirror symmetry $\mathcal{M}_x$. 
	
	Such a $\varphi $-junction exhibits Josephson diode effect in the under-damped regime, which can be seen through an analysis with the resistance-and-capacitance-shunted-junction (RCSJ) model. The RCSJ model converts the Josephson current problem into the classical motion of a mass point on a tilted Josephson potential, as shown in FIG. \ref{fig:1} (c), where the tilting term is $I_b \phi$, proportional to the bias current $I_b$. The critical currents $I_{c\pm} $ are the slopes of the tilting potential beyond which the mass point, with a static initial state, cannot stop moving and thus the average voltage $V\sim \langle d\phi/dt \rangle $ is nonzero. 	
	For a SC in a chosen macroscopic state, the heights of the closest barrier towards $+ \phi$ and $-\phi $ directions are different, yielding unequal critical values of the potential slopes. (The further barrier can be overcome by the accumulated kinetic energy in under-damped junctions.) Thus, the critical currents along opposite directions are unequal, i.e., $I_{c+}\neq I_{c-}$, realizing a zero-field JDE. 
	
	 The $\varphi$-junction and its resulting JDE bear resemblance to phenomena observed in twisted bilayers of $d$-wave superconductors \cite{yang2018PRB,can2021NP, zhao2023Science, volkov2024PRB}, but the underlying mechanisms differ significantly. In twisted d-wave SC bilayers, phase interference between the layers suppresses first-order Josephson tunneling, making second-order terms significant; the interplay of these harmonics leads to degenerate double minima in the free energy and thus a $\varphi$-junction with JDE. In contrast, our mechanism relies on strong Coulomb repulsion in the junction region. When $U_i=0$, Kramers degeneracy ensures an even number of electrons in the normal ground state for generic $\mu$, resulting in a conventional Josephson junction with $\phi_0=0$. Only when the Coulomb interaction is sufficiently strong can the $\varphi$-junction and the associated JDE emerge.
			
\begin{figure}
	\begin{center}
		\includegraphics[width=1\columnwidth]{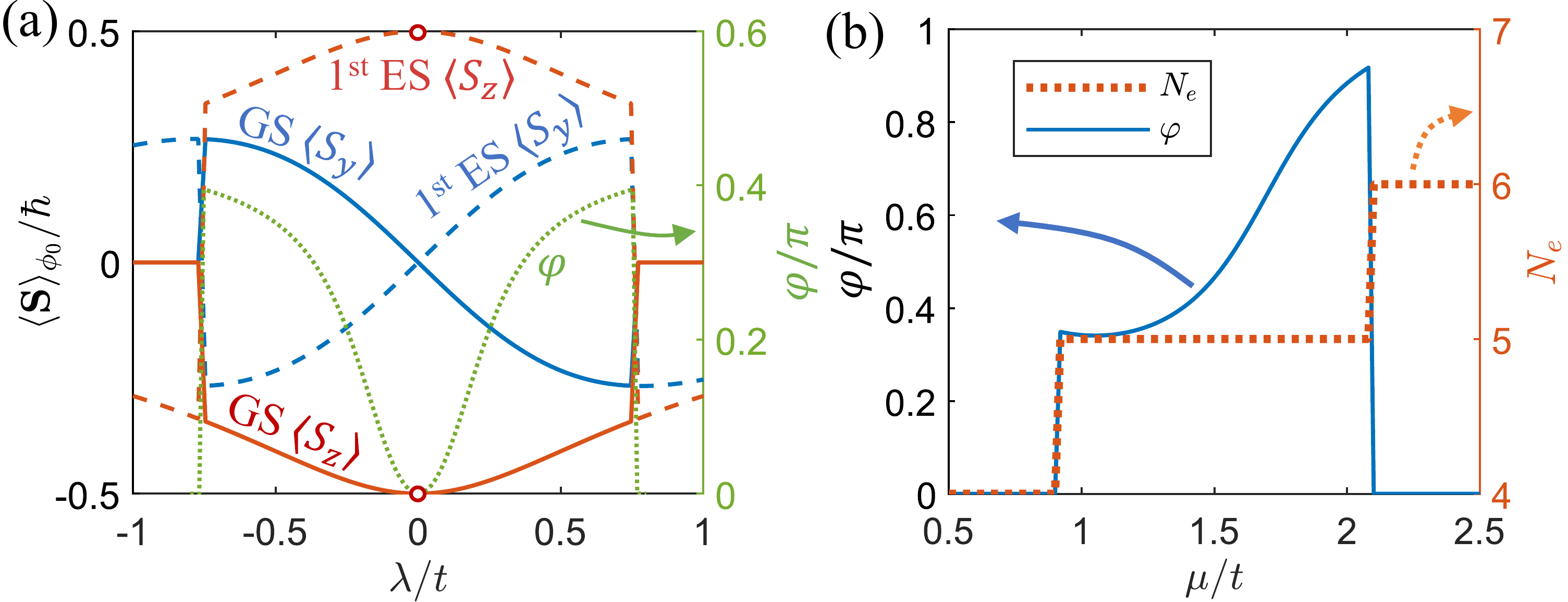} 
		\caption{(a) The ground-state Josephson phase $\varphi$ (right axis) and the spin polarization of the GS and the 1st ES as functions of the SOC strength. The open circles at $\lambda=0$ indicate the non-polarization due to degeneracy. (b) Variation of $\varphi$ and the total electron number $N_e$ with the chemical potential $\mu$. }
		\label{fig:2}
	\end{center}
	\end{figure}

	\paragraph{Symmetry requirements --- } 
	Besides the condition of odd $N_e$, certain symmetries must be broken to obtain a $\varphi$-junction. Breaking SU(2) spin-rotation symmetry is essential since the GS is always spin-degenerate for arbitrary $\phi$, i.e., $E_0(\phi)=E_1(\phi)$, if the SU(2) symmetry is preserved. 
	Then, the Josephson junction is almost conventional, except that a $\pi$-junction may emerge \footnote{Z. Zhang and J. J. He, to be published.}. $\mathcal{T}$  remains preserved and there is  no nonreciprocity. A Rashba term is introduced in Eq. (\ref{eq:1}) to break SU(2) and the dependence of $\varphi$ on the SOC strength is shown in FIG. \ref{fig:2} (a). 
    $\varphi(\lambda<0)$ shows an increase of $\varphi $ as $\lambda$ grows, with the absence of a linear dependence at small $\lambda$, indicating that the quantum state at $\varphi$ is insensitive to the sign of $\lambda$.  As $\lambda$ becomes large, the total number of electrons $N_e$ can turn even  and $\varphi$-junction collapse, i.e., $\varphi=0$. This happens also when $\mu$ is tuned, as shown in FIG. \ref{fig:2} (b). 
	The form of  SOC is not important neither, and replacing Eq. (\ref{eq:2}) with other forms of SOC does not affect the results significantly. Particularly, one may carefully construct spin-flipping hopping terms without breaking neither $\mathcal{T}$ or the spatial inversion symmetry $\mathcal{P}$, and the $\varphi$-junction can also be realized \cite{supp}. Thus, the SOC term is  needed only to break  SU(2) instead of $\mathcal{P}$, in sharp contrast to previously studied nonreciprocal transport in polar electronic systems where the sign of the SOC directly determines the sign of nonreciprocity. 
	

	\begin{table}
		\caption{Symmetry operations at fixed Josephson phase $\phi$. 
		}
		\label{tab:1}
		\begin{ruledtabular}
			\begin{tabular}{lcccc}
				& GSD & $S_x $&  $S_y $  & $S_z $\\
				\colrule
				SU(2) & \checkmark &   &   &  \\
				$\mathcal{M}_y$ &\checkmark &$ -1$ & $1$ & $-1$\\
				$\mathcal{M}_z$ & \checkmark &$ -1 $&$ -1$ & $1$ \\
				$\mathcal{C}_2\mathcal{T}$  & \checkmark & $1$ & $1$ & $-1$\\
				$\mathcal{P T}$  & \checkmark & $-1 $&$ -1$ & $-1$ \\
				$\mathcal{M }_x\mathcal{T}$  & $\times$ & $-1 $ &$ 1$ & $1 $
			\end{tabular}
		\end{ruledtabular}
	\end{table}
	
	Symmetries other than SU(2) that lead to $E_0(\phi)=E_1(\phi)$ include the mirror symmetries $\mathcal{M}_y$, $\mathcal{M}_z$, the two-fold rotation about the $z$-axis combined with the TR symmetry  $\mathcal{C}_2 \mathcal{T}$, and $\mathcal{PT}$.  The reason that $\mathcal{C}_2 \mathcal{T}$ and $\mathcal{PT}$ lead to degeneracy is readily seen  noting that both $\mathcal{C}_2$ (or $\mathcal{P}$) and $ \mathcal{T}$ reverse the phase $\phi \rightarrow -\phi$ and the combined symmetry squares to $-1$. Thus, the effect of this symmetry is similar to $\mathcal{T}$ except that $\phi$ remains unchanged. On the other hand, although $\mathcal{M}_{y/z}$ keep $\phi$ invariant, they square to one instead of $-1$. To see why they lead to degeneracy, we must take a closer look at the GS. 
	The Coulomb interaction leads to antiferromagnetic spin correlation.  Due to the odd number of electrons, the total spin of the GS is $|\bm S|=1/2$ (taking $\hbar=1$ throughout the manuscript), which is not altered when a Josephson phase is present as long as $U$ is large compared to the SC order parameter $\Delta$. 
	If a mirror symmetry, say $\mathcal{M}_y$, is present, let us choose the basis so that the GS can be labeled as $|\Psi_\pm  \rangle $  where $\pm $ denotes the spin components $S_{\pm}$ along an axis in the plane, say the $x$-axis.  
	Since $\mathcal{M}_y^{-1} S_\pm \mathcal{M}_y = -S_\mp$, one has $\mathcal{M}_y |\Psi_+ \rangle = e^{i\theta} |\Psi_-\rangle$ where an unimportant phase factor is included. Thus, the $|\Psi_\pm \rangle $ form the basis of a two-dimensional representation of the symmetry group, resulting in a two-fold degeneracy. To see this explicitly, one can combine $\mathcal{M}_y H(\phi)=H(\phi) \mathcal{M}_y$ with the above relation and get $H |\Psi_- \rangle = H \mathcal{M}_y|\Psi_+ \rangle e^{-i\theta} = \mathcal{M}_y H |\Psi_+ \rangle e^{-i\theta} = E_+ \mathcal{M}_y |\Psi_+ \rangle e^{-i\theta} =E_+ |\Psi_- \rangle $. Comparison with $H |\Psi_- \rangle = E_-|\Psi_- \rangle $ proves $E_+(\phi)=E_-(\phi)$. A similar analysis can be done for $\mathcal{M}_z$. By setting a non-uniform Coulomb interaction as shown in FIG. \ref{fig:1} (a), all the above symmetries that lead to ground-state degeneracy, as shown in TABLE \ref{tab:1}, are broken, and a $\varphi$-junction is achieved. Symmetry breaking induced by other parameters, such as a non-uniform $\mu$, is equally effective.

	\paragraph{Spin polarization --- }
	In the $\varphi$-junction here, the non-degenerate states are spin-polarized. FIG. \ref{fig:1} (d) shows the spin expectation values of the GS and the 1st ES as functions of $\phi$. The spin is polarized in the $y$-$z$ plane since  $\langle S_x \rangle =0 $ constantly due to the presence of the combined  symmetry $\mathcal{M}_x \mathcal{T}$ for arbitrary $\phi$, which flips $S_x$, as shown in TABLE \ref{tab:1}. This symmetry is present only for the specific geometry in FiG. \ref{fig:1} (a), and $\langle S_x \rangle $ is nonzero in other cases with broken $\mathcal{M}_x$ \cite{supp}.	
	The Josephson phase does not affect the spin-polarization of individual states. However, it varies the relative energy between the lowest two levels, as shown in FIG. \ref{fig:1} (b). Particularly, it interchanges the GS and the 1st ES across the degenerate points $\phi=0$ and $\phi=\pi$. This is expected since the Hamiltonian (\ref{eq:1}) respects $\mathcal{T}$ which reverses both $\phi$ and $\bm S$, dictating $\bm S(-\phi)=-\bm S(\phi)$. 

	Since a nonzero $\phi$ usually corresponds to a Josephson current, the spin-polarization at finite $\phi$ may remind the reader of a supercurrent Edelstein effect (SEE) found in spin-orbit coupled superconductors \cite{Edelstein1995PRL,he2019CP}. The $\varphi$-junction state have several key differences from the SEE. Firstly, the current at $\pm \varphi$ is zero, and thus the spin-polarization can happen without a bias current. Another difference lies in the role of SOC. 
	In the SEE, the polarity of the system is controlled by the SOC and the magnetization grows continuously as  $\lambda $ increases. The $\lambda$-dependence of $\langle S_y \rangle $ here is similar, but $\langle S_z (\lambda) \rangle $  remains finite as $\lambda \rightarrow 0^\pm $, as shown in FIG. \ref{fig:2} (a). For a small $\lambda$, the spin polarization has most prominent component along $z$-direction and  tilted towards $y$-axis as $\lambda$ increases, indicating that the SOC may alter the polarization but it does not directly polarize the system. In other words, the SOC mainly plays its role through breaking SU(2) instead of $\mathcal{P}$, in sharp contrast to the case of SEE. If $\lambda$ is so large that $N_e$ becomes an even integer, the states become degenerate and unpolarized. The polarization of this region shown in FIG. \ref{fig:2} (a) is caused by imperfect degeneracy due to numerical error. 
	
	\begin{figure}
		\includegraphics[width=1\columnwidth]{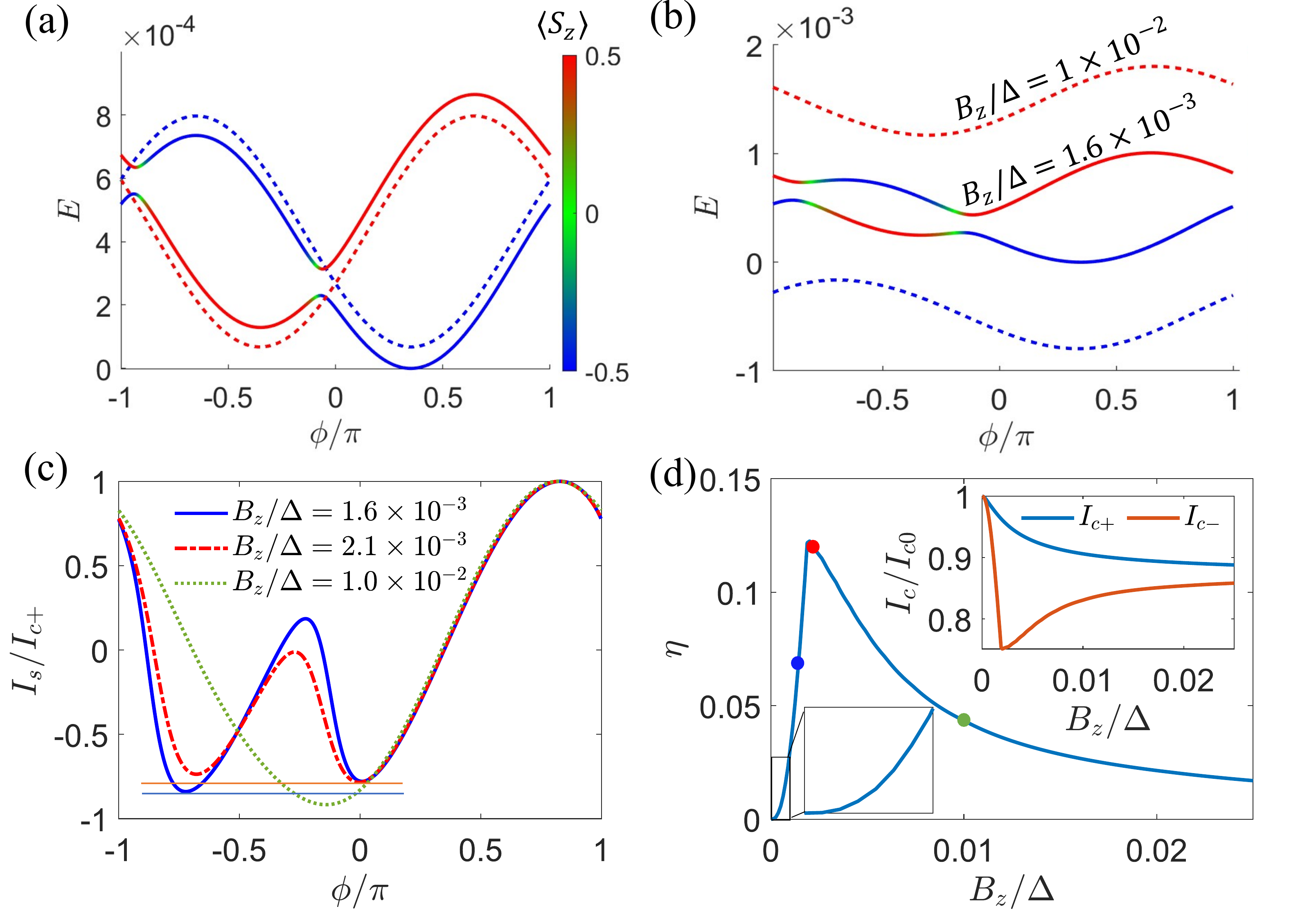} 
		\caption{ (a) Energy spectra under a small external Zeeman field $B_z=8\times 10^{-5}$ (solid) compared with those without field (dashed).  The color denotes spin-polarization along $z$-direction $\langle S_z \rangle $. (b) Energy spectra under $B_z=1.6 \times 10^{-4}$ (solid) and $B_z=1 \times 10^{-3}$ (dashed). The color bar is the same as that of (a). (c) Current phase relation with various $B_z$  marked in (d). The horizontal lines at the minima are guide for the eyes.  (d) The JDE  efficiency $\eta \equiv (I_{c+}-I_{c-})/(I_{c+}+I_{c-})$ and critical currents $I_{c\pm}$ (upper-right inset) as functions of $B_z$. The zoom-in at lower-left inset shows the nonlinear dependence at small field. }
		\label{fig:3}
	\end{figure}

\begin{figure}
		\includegraphics[width=0.7\columnwidth]{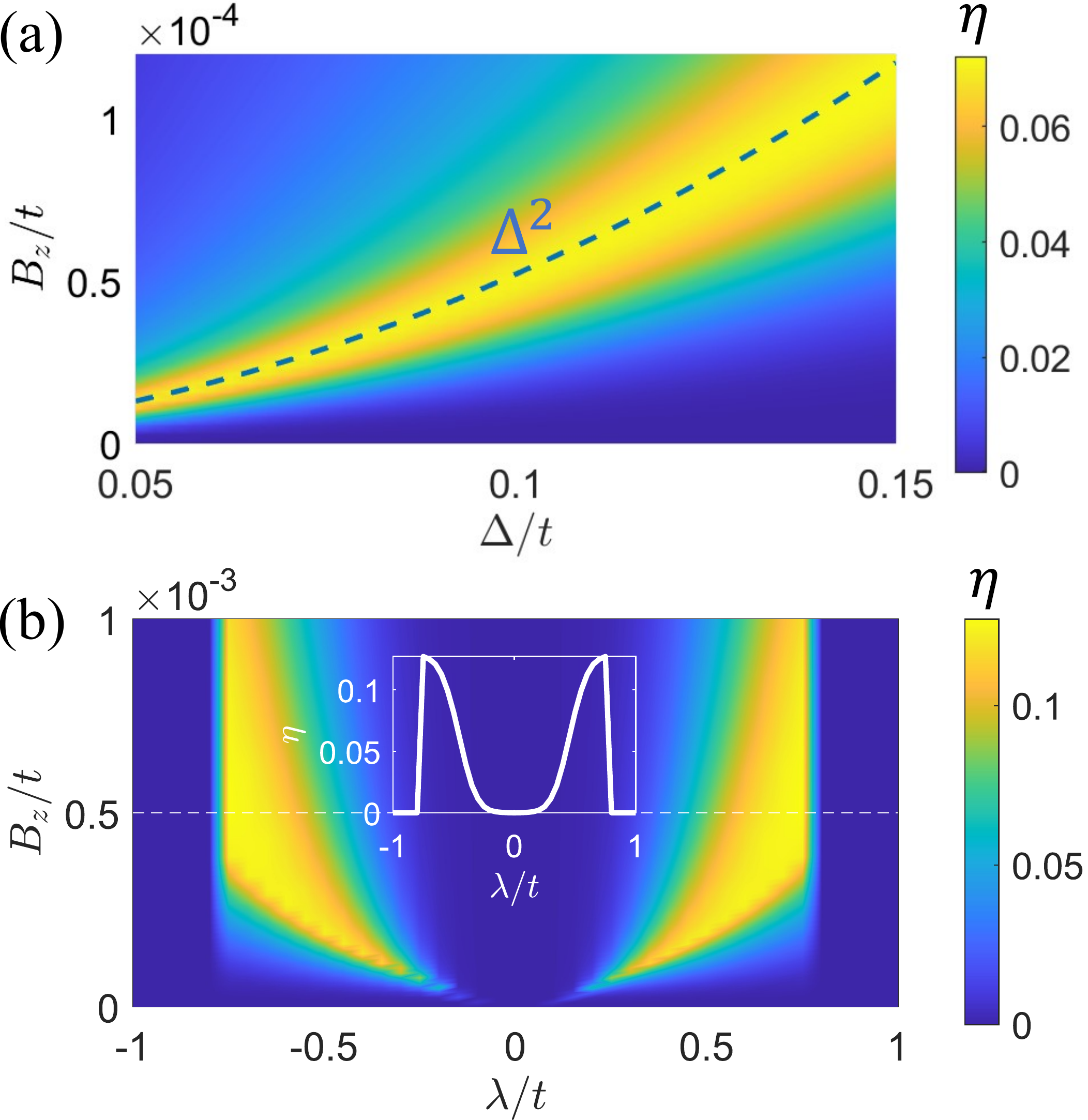} 
		\caption{ (a) A heatmap of the JDE efficiency $\eta$ as a function of the SC order parameter $\Delta$ and the Zeeman field $B_z$. The dashed curve is  $\Delta^2$ up to a constant scaling, showing that the peak value of the external field $B_p \sim \Delta^2$.  (b) $\eta$ as functions of the SOC $\lambda $ and $B_z$. The inset shows the line cut at $B_z/t=5\times 10^{-4}$. } 
		\label{fig:4}
	\end{figure}

	\paragraph{Nonzero magnetic field ---} The spontaneous-symmetry-broken GS of a $\varphi$-junction leads to JDE that may flip sign after superconducting-normal-superconducting transitions, because the GS is randomly chosen between $\pm \varphi$ each time the SC state nucleates \cite{zhao2023Science}.  On the other hand, a controllable JDE can be achieved in the over-damped regime by applying an magnetic field, which introduces a Zeeman term  	$		H_Z = -\sum_i \bm B \cdot \bm \sigma_{\alpha \beta }c^\dagger_{i\alpha} c_{j\beta}. 	$	The energy spectra for a small field $B_z$ along $z$-direction 	is shown in FIG. \ref{fig:3} (a) where the color denotes $\langle S_z \rangle$.  
	GSD at $\phi=0$ and $\pi$ are lifted due to the $\mathcal{T}$-breaking. By comparing to the zero-field spectrum, it is clear that the Zeeman field moves the energy of the state with $\langle S_z \rangle >0$  upward and that with $\langle S_z \rangle<0$  downward, as expected. 	
	Since $\langle \bm S \rangle $ form a nonzero angle with $\hat{\bm z}$,  a small $B_z$ introduces a coupling between the two states and opens gaps at the ``band-inversion" points. Such a coupling is manifested in the smooth transitions of the spin polarization as $\phi$ is varied, as shown by the color in FIG. \ref{fig:3} (a). Due to the band inversion, there are two local maxima and two local minima in the current-phase relation $I_s(\phi)$, as well as in $E_0(\phi)$. 
	As $B_z$ increases, two band-inversion points move closer and the coupling becomes stronger, as shown in FIG. \ref{fig:3} (b). The two minima in $I_s(\phi)$ get closer. Across a critical value, $B_p$, their relative height changes sign and the global minimum, i.e., $ I_{\min } \equiv  -I_{c-} $, transitions from one local minimum to the other, as shown in FIG. \ref{fig:3} (c). The diode efficiency $\eta \equiv (I_{c+}-I_{c-})/(I_{c+}+I_{c-}) $ reaches its maximum at the transition point, and further increasing $B_z$ suppresses $\eta$, as shown in FIG. \ref{fig:3} (d). 	
	When $B_z$ becomes larger than $E_J$, the energy difference between the two bands is large and there is no band-inversion, as shown by the dashed curves in FIG. \ref{fig:3} (b). The locking of spin by the external field is so strong that the Josephson phase cannot flip it. In this regime, the JDE becomes weak. 		
	A heatmap of $\eta$ as a function of the Zeeman field and the pairing order parameter is shown in FIG. \ref{fig:4} (a). The peak position follows the relation $B_p\sim \Delta^2 $,  consistent with the  analysis above since the ``band width" is roughly $ E_J \sim \Delta^2$. 

	A Zeeman field in cooperation with SOC may induce SDE in non-interacting junctions as well. However, it requires a rather large field to obtain sizable $\eta$, while a tiny field (smaller than the SC gap by at least two orders of magnitude) causes a JDE of more than $10\%$ in the strongly correlated JJ. 
	The Zeeman field here triggered the polarization and the JDE very efficiently due to the enhancement from strong magnetic correlations. Importantly, the SOC does not determine the diode polarity --- the JDE efficiency $\eta$ is an even function of the spin-orbit strength, consistent with the behavior of $\langle S_z \rangle $ shown in FIG. \ref{fig:2} (a).

	\paragraph{Discussion ---} 
	
	In summary, we studied a Josephson junction through a region with strong electron-electron interaction and revealed a new mechanism of JDE in which the strong correlation plays the key role. It relies on the realization of a $\varphi$-junction by spontaneous time-reversal symmetry breaking, which is possible when the total electron number in the junction is odd and symmetry requirements are met. The effect of external Zeeman field is further discussed and a non-monotonic dependence of the JDE on the field is found which peak at a tiny field compared to other energy scales. 
	
	This study is carried out at zero temperature $T=0$. When $T\neq 0$, the upper energy levels will get involved and the free energy becomes a smooth function of $\phi$ even at zero field, unlike $E_0(\phi)$ in FIG. \ref{fig:1} (b) \cite{supp}. As long as $T \ll \Delta $, the main conclusions here remain valid. For a larger $T$, the $\phi$-junction will become a conventional one, but the JDE peaking at a tiny external field must remain since the strong correlation survives. 
	
	Besides explicit Josephson junctions, our theory may also be relevant to (quasi) two-dimensional material systems where correlation is strong and inhomogeneity may exist so that the system separates into weakly linked SC islands. Multiple junctions may coexist and the junction geometries can be very different. However, the details of the junction only affect the results quantitatively as long as the electron parity is odd and symmetries that lead to $E_0(\phi)=E_1(\pi)$ are broken \cite{supp}. Thus, our minimal model may capture the fundamental physics behind nonreciprocal supercurrents in such systems.

%
%
 
	\paragraph{Acknowledgment ---} 
	
	This work is supported by the National Science Foundation of China (Grant No. 12488101) and the Quantum Science and Technology-National Science and Technology Major Project (Grant No. 2021ZD0302800).

	\bibliography{Ref_SDE}	

\begin{thebibliography}{59}%
\makeatletter
\providecommand \@ifxundefined [1]{%
 \@ifx{#1\undefined}
}%
\providecommand \@ifnum [1]{%
 \ifnum #1\expandafter \@firstoftwo
 \else \expandafter \@secondoftwo
 \fi
}%
\providecommand \@ifx [1]{%
 \ifx #1\expandafter \@firstoftwo
 \else \expandafter \@secondoftwo
 \fi
}%
\providecommand \natexlab [1]{#1}%
\providecommand \enquote  [1]{``#1''}%
\providecommand \bibnamefont  [1]{#1}%
\providecommand \bibfnamefont [1]{#1}%
\providecommand \citenamefont [1]{#1}%
\providecommand \href@noop [0]{\@secondoftwo}%
\providecommand \href [0]{\begingroup \@sanitize@url \@href}%
\providecommand \@href[1]{\@@startlink{#1}\@@href}%
\providecommand \@@href[1]{\endgroup#1\@@endlink}%
\providecommand \@sanitize@url [0]{\catcode `\\12\catcode `\$12\catcode
  `\&12\catcode `\#12\catcode `\^12\catcode `\_12\catcode `\%12\relax}%
\providecommand \@@startlink[1]{}%
\providecommand \@@endlink[0]{}%
\providecommand \url  [0]{\begingroup\@sanitize@url \@url }%
\providecommand \@url [1]{\endgroup\@href {#1}{\urlprefix }}%
\providecommand \urlprefix  [0]{URL }%
\providecommand \Eprint [0]{\href }%
\providecommand \doibase [0]{https://doi.org/}%
\providecommand \selectlanguage [0]{\@gobble}%
\providecommand \bibinfo  [0]{\@secondoftwo}%
\providecommand \bibfield  [0]{\@secondoftwo}%
\providecommand \translation [1]{[#1]}%
\providecommand \BibitemOpen [0]{}%
\providecommand \bibitemStop [0]{}%
\providecommand \bibitemNoStop [0]{.\EOS\space}%
\providecommand \EOS [0]{\spacefactor3000\relax}%
\providecommand \BibitemShut  [1]{\csname bibitem#1\endcsname}%
\let\auto@bib@innerbib\@empty
\bibitem [{\citenamefont {Ando}\ \emph {et~al.}(2020)\citenamefont {Ando},
  \citenamefont {Miyasaka}, \citenamefont {Li}, \citenamefont {Ishizuka},
  \citenamefont {Arakawa}, \citenamefont {Shiota}, \citenamefont {Moriyama},
  \citenamefont {Yanase},\ and\ \citenamefont {Ono}}]{Ando_Nature2020}%
  \BibitemOpen
  \bibfield  {author} {\bibinfo {author} {\bibfnamefont {F.}~\bibnamefont
  {Ando}}, \bibinfo {author} {\bibfnamefont {Y.}~\bibnamefont {Miyasaka}},
  \bibinfo {author} {\bibfnamefont {T.}~\bibnamefont {Li}}, \bibinfo {author}
  {\bibfnamefont {J.}~\bibnamefont {Ishizuka}}, \bibinfo {author}
  {\bibfnamefont {T.}~\bibnamefont {Arakawa}}, \bibinfo {author} {\bibfnamefont
  {Y.}~\bibnamefont {Shiota}}, \bibinfo {author} {\bibfnamefont
  {T.}~\bibnamefont {Moriyama}}, \bibinfo {author} {\bibfnamefont
  {Y.}~\bibnamefont {Yanase}},\ and\ \bibinfo {author} {\bibfnamefont
  {T.}~\bibnamefont {Ono}},\ }\bibfield  {title} {\bibinfo {title} {Observation
  of superconducting diode effect},\ }\href
  {https://doi.org/10.1038/s41586-020-2590-4} {\bibfield  {journal} {\bibinfo
  {journal} {Nature}\ }\textbf {\bibinfo {volume} {584}},\ \bibinfo {pages}
  {373} (\bibinfo {year} {2020})}\BibitemShut {NoStop}%
\bibitem [{\citenamefont {Hou}\ \emph {et~al.}(2023{\natexlab{a}})\citenamefont
  {Hou}, \citenamefont {Nichele}, \citenamefont {Chi}, \citenamefont
  {Lodesani}, \citenamefont {Wu}, \citenamefont {Ritter}, \citenamefont
  {Haxell}, \citenamefont {Davydova}, \citenamefont {Ili{\'c}}, \citenamefont
  {Glezakou-Elbert} \emph {et~al.}}]{hou2023PRL}%
  \BibitemOpen
  \bibfield  {author} {\bibinfo {author} {\bibfnamefont {Y.}~\bibnamefont
  {Hou}}, \bibinfo {author} {\bibfnamefont {F.}~\bibnamefont {Nichele}},
  \bibinfo {author} {\bibfnamefont {H.}~\bibnamefont {Chi}}, \bibinfo {author}
  {\bibfnamefont {A.}~\bibnamefont {Lodesani}}, \bibinfo {author}
  {\bibfnamefont {Y.}~\bibnamefont {Wu}}, \bibinfo {author} {\bibfnamefont
  {M.~F.}\ \bibnamefont {Ritter}}, \bibinfo {author} {\bibfnamefont {D.~Z.}\
  \bibnamefont {Haxell}}, \bibinfo {author} {\bibfnamefont {M.}~\bibnamefont
  {Davydova}}, \bibinfo {author} {\bibfnamefont {S.}~\bibnamefont {Ili{\'c}}},
  \bibinfo {author} {\bibfnamefont {O.}~\bibnamefont {Glezakou-Elbert}}, \emph
  {et~al.},\ }\bibfield  {title} {\bibinfo {title} {Ubiquitous superconducting
  diode effect in superconductor thin films},\ }\href@noop {} {\bibfield
  {journal} {\bibinfo  {journal} {Phys. Rev. Lett.}\ }\textbf {\bibinfo
  {volume} {131}},\ \bibinfo {pages} {027001} (\bibinfo {year}
  {2023}{\natexlab{a}})}\BibitemShut {NoStop}%
\bibitem [{\citenamefont {Gutfreund}\ \emph {et~al.}(2023)\citenamefont
  {Gutfreund}, \citenamefont {Matsuki}, \citenamefont {Plastovets},
  \citenamefont {Noah}, \citenamefont {Gorzawski}, \citenamefont {Fridman},
  \citenamefont {Yang}, \citenamefont {Buzdin}, \citenamefont {Millo},
  \citenamefont {Robinson} \emph {et~al.}}]{gutfreund2023NC}%
  \BibitemOpen
  \bibfield  {author} {\bibinfo {author} {\bibfnamefont {A.}~\bibnamefont
  {Gutfreund}}, \bibinfo {author} {\bibfnamefont {H.}~\bibnamefont {Matsuki}},
  \bibinfo {author} {\bibfnamefont {V.}~\bibnamefont {Plastovets}}, \bibinfo
  {author} {\bibfnamefont {A.}~\bibnamefont {Noah}}, \bibinfo {author}
  {\bibfnamefont {L.}~\bibnamefont {Gorzawski}}, \bibinfo {author}
  {\bibfnamefont {N.}~\bibnamefont {Fridman}}, \bibinfo {author} {\bibfnamefont
  {G.}~\bibnamefont {Yang}}, \bibinfo {author} {\bibfnamefont {A.}~\bibnamefont
  {Buzdin}}, \bibinfo {author} {\bibfnamefont {O.}~\bibnamefont {Millo}},
  \bibinfo {author} {\bibfnamefont {J.~W.}\ \bibnamefont {Robinson}}, \emph
  {et~al.},\ }\bibfield  {title} {\bibinfo {title} {Direct observation of a
  superconducting vortex diode},\ }\href@noop {} {\bibfield  {journal}
  {\bibinfo  {journal} {Nature Communications}\ }\textbf {\bibinfo {volume}
  {14}},\ \bibinfo {pages} {1630} (\bibinfo {year} {2023})}\BibitemShut
  {NoStop}%
\bibitem [{\citenamefont {Li}\ \emph {et~al.}(2024)\citenamefont {Li},
  \citenamefont {Yan}, \citenamefont {Hong}, \citenamefont {Sheng},
  \citenamefont {Wang}, \citenamefont {Dou}, \citenamefont {Guo}, \citenamefont
  {Shi}, \citenamefont {Su}, \citenamefont {Lyu} \emph {et~al.}}]{li2024NC}%
  \BibitemOpen
  \bibfield  {author} {\bibinfo {author} {\bibfnamefont {Y.}~\bibnamefont
  {Li}}, \bibinfo {author} {\bibfnamefont {D.}~\bibnamefont {Yan}}, \bibinfo
  {author} {\bibfnamefont {Y.}~\bibnamefont {Hong}}, \bibinfo {author}
  {\bibfnamefont {H.}~\bibnamefont {Sheng}}, \bibinfo {author} {\bibfnamefont
  {A.}~\bibnamefont {Wang}}, \bibinfo {author} {\bibfnamefont {Z.}~\bibnamefont
  {Dou}}, \bibinfo {author} {\bibfnamefont {X.}~\bibnamefont {Guo}}, \bibinfo
  {author} {\bibfnamefont {X.}~\bibnamefont {Shi}}, \bibinfo {author}
  {\bibfnamefont {Z.}~\bibnamefont {Su}}, \bibinfo {author} {\bibfnamefont
  {Z.}~\bibnamefont {Lyu}}, \emph {et~al.},\ }\bibfield  {title} {\bibinfo
  {title} {Interfering josephson diode effect in ta2pd3te5 asymmetric edge
  interferometer},\ }\href@noop {} {\bibfield  {journal} {\bibinfo  {journal}
  {Nature Communications}\ }\textbf {\bibinfo {volume} {15}},\ \bibinfo {pages}
  {9031} (\bibinfo {year} {2024})}\BibitemShut {NoStop}%
\bibitem [{\citenamefont {Baumgartner}\ \emph {et~al.}(2022)\citenamefont
  {Baumgartner}, \citenamefont {Fuchs}, \citenamefont {Costa}, \citenamefont
  {Reinhardt}, \citenamefont {Gronin}, \citenamefont {Gardner}, \citenamefont
  {Lindemann}, \citenamefont {Manfra}, \citenamefont {Faria~Junior},
  \citenamefont {Kochan} \emph {et~al.}}]{baumgartner2022NN}%
  \BibitemOpen
  \bibfield  {author} {\bibinfo {author} {\bibfnamefont {C.}~\bibnamefont
  {Baumgartner}}, \bibinfo {author} {\bibfnamefont {L.}~\bibnamefont {Fuchs}},
  \bibinfo {author} {\bibfnamefont {A.}~\bibnamefont {Costa}}, \bibinfo
  {author} {\bibfnamefont {S.}~\bibnamefont {Reinhardt}}, \bibinfo {author}
  {\bibfnamefont {S.}~\bibnamefont {Gronin}}, \bibinfo {author} {\bibfnamefont
  {G.~C.}\ \bibnamefont {Gardner}}, \bibinfo {author} {\bibfnamefont
  {T.}~\bibnamefont {Lindemann}}, \bibinfo {author} {\bibfnamefont {M.~J.}\
  \bibnamefont {Manfra}}, \bibinfo {author} {\bibfnamefont {P.~E.}\
  \bibnamefont {Faria~Junior}}, \bibinfo {author} {\bibfnamefont
  {D.}~\bibnamefont {Kochan}}, \emph {et~al.},\ }\bibfield  {title} {\bibinfo
  {title} {Supercurrent rectification and magnetochiral effects in symmetric
  josephson junctions},\ }\href@noop {} {\bibfield  {journal} {\bibinfo
  {journal} {Nature nanotechnology}\ }\textbf {\bibinfo {volume} {17}},\
  \bibinfo {pages} {39} (\bibinfo {year} {2022})}\BibitemShut {NoStop}%
\bibitem [{\citenamefont {Pal}\ \emph {et~al.}(2022)\citenamefont {Pal},
  \citenamefont {Chakraborty}, \citenamefont {Sivakumar}, \citenamefont
  {Davydova}, \citenamefont {Gopi}, \citenamefont {Pandeya}, \citenamefont
  {Krieger}, \citenamefont {Zhang}, \citenamefont {Date}, \citenamefont {Ju}
  \emph {et~al.}}]{pal2022NP}%
  \BibitemOpen
  \bibfield  {author} {\bibinfo {author} {\bibfnamefont {B.}~\bibnamefont
  {Pal}}, \bibinfo {author} {\bibfnamefont {A.}~\bibnamefont {Chakraborty}},
  \bibinfo {author} {\bibfnamefont {P.~K.}\ \bibnamefont {Sivakumar}}, \bibinfo
  {author} {\bibfnamefont {M.}~\bibnamefont {Davydova}}, \bibinfo {author}
  {\bibfnamefont {A.~K.}\ \bibnamefont {Gopi}}, \bibinfo {author}
  {\bibfnamefont {A.~K.}\ \bibnamefont {Pandeya}}, \bibinfo {author}
  {\bibfnamefont {J.~A.}\ \bibnamefont {Krieger}}, \bibinfo {author}
  {\bibfnamefont {Y.}~\bibnamefont {Zhang}}, \bibinfo {author} {\bibfnamefont
  {M.}~\bibnamefont {Date}}, \bibinfo {author} {\bibfnamefont {S.}~\bibnamefont
  {Ju}}, \emph {et~al.},\ }\bibfield  {title} {\bibinfo {title} {Josephson
  diode effect from cooper pair momentum in a topological semimetal},\
  }\href@noop {} {\bibfield  {journal} {\bibinfo  {journal} {Nature physics}\
  }\textbf {\bibinfo {volume} {18}},\ \bibinfo {pages} {1228} (\bibinfo {year}
  {2022})}\BibitemShut {NoStop}%
\bibitem [{\citenamefont {Bauriedl}\ \emph {et~al.}(2022)\citenamefont
  {Bauriedl}, \citenamefont {B{\"a}uml}, \citenamefont {Fuchs}, \citenamefont
  {Baumgartner}, \citenamefont {Paulik}, \citenamefont {Bauer}, \citenamefont
  {Lin}, \citenamefont {Lupton}, \citenamefont {Taniguchi}, \citenamefont
  {Watanabe} \emph {et~al.}}]{bauriedl2022NC}%
  \BibitemOpen
  \bibfield  {author} {\bibinfo {author} {\bibfnamefont {L.}~\bibnamefont
  {Bauriedl}}, \bibinfo {author} {\bibfnamefont {C.}~\bibnamefont {B{\"a}uml}},
  \bibinfo {author} {\bibfnamefont {L.}~\bibnamefont {Fuchs}}, \bibinfo
  {author} {\bibfnamefont {C.}~\bibnamefont {Baumgartner}}, \bibinfo {author}
  {\bibfnamefont {N.}~\bibnamefont {Paulik}}, \bibinfo {author} {\bibfnamefont
  {J.~M.}\ \bibnamefont {Bauer}}, \bibinfo {author} {\bibfnamefont {K.-Q.}\
  \bibnamefont {Lin}}, \bibinfo {author} {\bibfnamefont {J.~M.}\ \bibnamefont
  {Lupton}}, \bibinfo {author} {\bibfnamefont {T.}~\bibnamefont {Taniguchi}},
  \bibinfo {author} {\bibfnamefont {K.}~\bibnamefont {Watanabe}}, \emph
  {et~al.},\ }\bibfield  {title} {\bibinfo {title} {Supercurrent diode effect
  and magnetochiral anisotropy in few-layer nbse2},\ }\href@noop {} {\bibfield
  {journal} {\bibinfo  {journal} {Nature communications}\ }\textbf {\bibinfo
  {volume} {13}},\ \bibinfo {pages} {4266} (\bibinfo {year}
  {2022})}\BibitemShut {NoStop}%
\bibitem [{\citenamefont {Chen}\ \emph {et~al.}(2024)\citenamefont {Chen},
  \citenamefont {Wang}, \citenamefont {Ye}, \citenamefont {Wang}, \citenamefont
  {Zhou}, \citenamefont {Tang}, \citenamefont {Wang}, \citenamefont {Wang},
  \citenamefont {Zhang}, \citenamefont {Mei} \emph {et~al.}}]{chen2024AFM}%
  \BibitemOpen
  \bibfield  {author} {\bibinfo {author} {\bibfnamefont {P.}~\bibnamefont
  {Chen}}, \bibinfo {author} {\bibfnamefont {G.}~\bibnamefont {Wang}}, \bibinfo
  {author} {\bibfnamefont {B.}~\bibnamefont {Ye}}, \bibinfo {author}
  {\bibfnamefont {J.}~\bibnamefont {Wang}}, \bibinfo {author} {\bibfnamefont
  {L.}~\bibnamefont {Zhou}}, \bibinfo {author} {\bibfnamefont {Z.}~\bibnamefont
  {Tang}}, \bibinfo {author} {\bibfnamefont {L.}~\bibnamefont {Wang}}, \bibinfo
  {author} {\bibfnamefont {J.}~\bibnamefont {Wang}}, \bibinfo {author}
  {\bibfnamefont {W.}~\bibnamefont {Zhang}}, \bibinfo {author} {\bibfnamefont
  {J.}~\bibnamefont {Mei}}, \emph {et~al.},\ }\bibfield  {title} {\bibinfo
  {title} {Edelstein effect induced superconducting diode effect in inversion
  symmetry breaking mote2 josephson junctions},\ }\href@noop {} {\bibfield
  {journal} {\bibinfo  {journal} {Advanced Functional Materials}\ }\textbf
  {\bibinfo {volume} {34}},\ \bibinfo {pages} {2311229} (\bibinfo {year}
  {2024})}\BibitemShut {NoStop}%
\bibitem [{\citenamefont {Kim}\ \emph {et~al.}(2024)\citenamefont {Kim},
  \citenamefont {Jeon}, \citenamefont {Sivakumar}, \citenamefont {Jeon},
  \citenamefont {Koerner}, \citenamefont {Woltersdorf},\ and\ \citenamefont
  {Parkin}}]{kim2024NC}%
  \BibitemOpen
  \bibfield  {author} {\bibinfo {author} {\bibfnamefont {J.-K.}\ \bibnamefont
  {Kim}}, \bibinfo {author} {\bibfnamefont {K.-R.}\ \bibnamefont {Jeon}},
  \bibinfo {author} {\bibfnamefont {P.~K.}\ \bibnamefont {Sivakumar}}, \bibinfo
  {author} {\bibfnamefont {J.}~\bibnamefont {Jeon}}, \bibinfo {author}
  {\bibfnamefont {C.}~\bibnamefont {Koerner}}, \bibinfo {author} {\bibfnamefont
  {G.}~\bibnamefont {Woltersdorf}},\ and\ \bibinfo {author} {\bibfnamefont
  {S.~S.}\ \bibnamefont {Parkin}},\ }\bibfield  {title} {\bibinfo {title}
  {Intrinsic supercurrent non-reciprocity coupled to the crystal structure of a
  van der waals josephson barrier},\ }\href@noop {} {\bibfield  {journal}
  {\bibinfo  {journal} {Nature Communications}\ }\textbf {\bibinfo {volume}
  {15}},\ \bibinfo {pages} {1120} (\bibinfo {year} {2024})}\BibitemShut
  {NoStop}%
\bibitem [{\citenamefont {Guan}\ \emph {et~al.}(2026)\citenamefont {Guan},
  \citenamefont {Yan}, \citenamefont {Zhang}, \citenamefont {Sun},
  \citenamefont {Chen}, \citenamefont {Zhao}, \citenamefont {Zhao},
  \citenamefont {Gao}, \citenamefont {He},\ and\ \citenamefont
  {Wang}}]{guan2026CP}%
  \BibitemOpen
  \bibfield  {author} {\bibinfo {author} {\bibfnamefont {H.}~\bibnamefont
  {Guan}}, \bibinfo {author} {\bibfnamefont {C.}~\bibnamefont {Yan}}, \bibinfo
  {author} {\bibfnamefont {Z.}~\bibnamefont {Zhang}}, \bibinfo {author}
  {\bibfnamefont {Y.}~\bibnamefont {Sun}}, \bibinfo {author} {\bibfnamefont
  {Q.}~\bibnamefont {Chen}}, \bibinfo {author} {\bibfnamefont {X.}~\bibnamefont
  {Zhao}}, \bibinfo {author} {\bibfnamefont {C.}~\bibnamefont {Zhao}}, \bibinfo
  {author} {\bibfnamefont {B.}~\bibnamefont {Gao}}, \bibinfo {author}
  {\bibfnamefont {J.~J.}\ \bibnamefont {He}},\ and\ \bibinfo {author}
  {\bibfnamefont {S.}~\bibnamefont {Wang}},\ }\bibfield  {title} {\bibinfo
  {title} {Dual-mode superconducting diode effect enabled by in-plane and
  out-of-plane magnetic field},\ }\href@noop {} {\bibfield  {journal} {\bibinfo
   {journal} {Communications Physics}\ } (\bibinfo {year} {2026})}\BibitemShut
  {NoStop}%
\bibitem [{\citenamefont {Wu}\ \emph {et~al.}(2022)\citenamefont {Wu},
  \citenamefont {Wang}, \citenamefont {Xu}, \citenamefont {Sivakumar},
  \citenamefont {Pasco}, \citenamefont {Filippozzi}, \citenamefont {Parkin},
  \citenamefont {Zeng}, \citenamefont {McQueen},\ and\ \citenamefont
  {Ali}}]{Wu_Nature2022}%
  \BibitemOpen
  \bibfield  {author} {\bibinfo {author} {\bibfnamefont {H.}~\bibnamefont
  {Wu}}, \bibinfo {author} {\bibfnamefont {Y.}~\bibnamefont {Wang}}, \bibinfo
  {author} {\bibfnamefont {Y.}~\bibnamefont {Xu}}, \bibinfo {author}
  {\bibfnamefont {P.~K.}\ \bibnamefont {Sivakumar}}, \bibinfo {author}
  {\bibfnamefont {C.}~\bibnamefont {Pasco}}, \bibinfo {author} {\bibfnamefont
  {U.}~\bibnamefont {Filippozzi}}, \bibinfo {author} {\bibfnamefont {S.~S.~P.}\
  \bibnamefont {Parkin}}, \bibinfo {author} {\bibfnamefont {Y.-J.}\
  \bibnamefont {Zeng}}, \bibinfo {author} {\bibfnamefont {T.}~\bibnamefont
  {McQueen}},\ and\ \bibinfo {author} {\bibfnamefont {M.~N.}\ \bibnamefont
  {Ali}},\ }\bibfield  {title} {\bibinfo {title} {The field-free josephson
  diode in a van der waals heterostructure},\ }\href
  {https://doi.org/10.1038/s41586-022-04504-8} {\bibfield  {journal} {\bibinfo
  {journal} {Nature}\ }\textbf {\bibinfo {volume} {604}},\ \bibinfo {pages}
  {653} (\bibinfo {year} {2022})}\BibitemShut {NoStop}%
\bibitem [{\citenamefont {Liu}\ \emph {et~al.}(2024)\citenamefont {Liu},
  \citenamefont {Itahashi}, \citenamefont {Aoki}, \citenamefont {Dong},
  \citenamefont {Wang}, \citenamefont {Ogawa}, \citenamefont {Ideue},\ and\
  \citenamefont {Iwasa}}]{liu2024SA}%
  \BibitemOpen
  \bibfield  {author} {\bibinfo {author} {\bibfnamefont {F.}~\bibnamefont
  {Liu}}, \bibinfo {author} {\bibfnamefont {Y.~M.}\ \bibnamefont {Itahashi}},
  \bibinfo {author} {\bibfnamefont {S.}~\bibnamefont {Aoki}}, \bibinfo {author}
  {\bibfnamefont {Y.}~\bibnamefont {Dong}}, \bibinfo {author} {\bibfnamefont
  {Z.}~\bibnamefont {Wang}}, \bibinfo {author} {\bibfnamefont {N.}~\bibnamefont
  {Ogawa}}, \bibinfo {author} {\bibfnamefont {T.}~\bibnamefont {Ideue}},\ and\
  \bibinfo {author} {\bibfnamefont {Y.}~\bibnamefont {Iwasa}},\ }\bibfield
  {title} {\bibinfo {title} {Superconducting diode effect under time-reversal
  symmetry},\ }\href@noop {} {\bibfield  {journal} {\bibinfo  {journal}
  {Science Advances}\ }\textbf {\bibinfo {volume} {10}},\ \bibinfo {pages}
  {eado1502} (\bibinfo {year} {2024})}\BibitemShut {NoStop}%
\bibitem [{\citenamefont {Ma}\ \emph {et~al.}(2025{\natexlab{a}})\citenamefont
  {Ma}, \citenamefont {Wang}, \citenamefont {Zhuo}, \citenamefont {Lei},
  \citenamefont {Wang}, \citenamefont {Wang}, \citenamefont {Chen},
  \citenamefont {Wang}, \citenamefont {Ge}, \citenamefont {Wang} \emph
  {et~al.}}]{ma2025CP}%
  \BibitemOpen
  \bibfield  {author} {\bibinfo {author} {\bibfnamefont {J.}~\bibnamefont
  {Ma}}, \bibinfo {author} {\bibfnamefont {H.}~\bibnamefont {Wang}}, \bibinfo
  {author} {\bibfnamefont {W.}~\bibnamefont {Zhuo}}, \bibinfo {author}
  {\bibfnamefont {B.}~\bibnamefont {Lei}}, \bibinfo {author} {\bibfnamefont
  {S.}~\bibnamefont {Wang}}, \bibinfo {author} {\bibfnamefont {W.}~\bibnamefont
  {Wang}}, \bibinfo {author} {\bibfnamefont {X.-Y.}\ \bibnamefont {Chen}},
  \bibinfo {author} {\bibfnamefont {Z.-Y.}\ \bibnamefont {Wang}}, \bibinfo
  {author} {\bibfnamefont {B.}~\bibnamefont {Ge}}, \bibinfo {author}
  {\bibfnamefont {Z.}~\bibnamefont {Wang}}, \emph {et~al.},\ }\bibfield
  {title} {\bibinfo {title} {Field-free josephson diode effect in nbse2 van der
  waals junction},\ }\href@noop {} {\bibfield  {journal} {\bibinfo  {journal}
  {Communications Physics}\ }\textbf {\bibinfo {volume} {8}},\ \bibinfo {pages}
  {125} (\bibinfo {year} {2025}{\natexlab{a}})}\BibitemShut {NoStop}%
\bibitem [{\citenamefont {Anwar}\ \emph {et~al.}(2023)\citenamefont {Anwar},
  \citenamefont {Nakamura}, \citenamefont {Ishiguro}, \citenamefont {Arif},
  \citenamefont {Robinson}, \citenamefont {Yonezawa}, \citenamefont {Sigrist},\
  and\ \citenamefont {Maeno}}]{anwar2023CP}%
  \BibitemOpen
  \bibfield  {author} {\bibinfo {author} {\bibfnamefont {M.~S.}\ \bibnamefont
  {Anwar}}, \bibinfo {author} {\bibfnamefont {T.}~\bibnamefont {Nakamura}},
  \bibinfo {author} {\bibfnamefont {R.}~\bibnamefont {Ishiguro}}, \bibinfo
  {author} {\bibfnamefont {S.}~\bibnamefont {Arif}}, \bibinfo {author}
  {\bibfnamefont {J.~W.}\ \bibnamefont {Robinson}}, \bibinfo {author}
  {\bibfnamefont {S.}~\bibnamefont {Yonezawa}}, \bibinfo {author}
  {\bibfnamefont {M.}~\bibnamefont {Sigrist}},\ and\ \bibinfo {author}
  {\bibfnamefont {Y.}~\bibnamefont {Maeno}},\ }\bibfield  {title} {\bibinfo
  {title} {Spontaneous superconducting diode effect in non-magnetic
  nb/ru/sr2ruo4 topological junctions},\ }\href@noop {} {\bibfield  {journal}
  {\bibinfo  {journal} {Communications Physics}\ }\textbf {\bibinfo {volume}
  {6}},\ \bibinfo {pages} {290} (\bibinfo {year} {2023})}\BibitemShut {NoStop}%
\bibitem [{\citenamefont {Diez-Merida}\ \emph {et~al.}(2023)\citenamefont
  {Diez-Merida}, \citenamefont {D{\'\i}ez-Carl{\'o}n}, \citenamefont {Yang},
  \citenamefont {Xie}, \citenamefont {Gao}, \citenamefont {Senior},
  \citenamefont {Watanabe}, \citenamefont {Taniguchi}, \citenamefont {Lu},
  \citenamefont {Higginbotham} \emph {et~al.}}]{diez2023NC}%
  \BibitemOpen
  \bibfield  {author} {\bibinfo {author} {\bibfnamefont {J.}~\bibnamefont
  {Diez-Merida}}, \bibinfo {author} {\bibfnamefont {A.}~\bibnamefont
  {D{\'\i}ez-Carl{\'o}n}}, \bibinfo {author} {\bibfnamefont {S.}~\bibnamefont
  {Yang}}, \bibinfo {author} {\bibfnamefont {Y.-M.}\ \bibnamefont {Xie}},
  \bibinfo {author} {\bibfnamefont {X.-J.}\ \bibnamefont {Gao}}, \bibinfo
  {author} {\bibfnamefont {J.}~\bibnamefont {Senior}}, \bibinfo {author}
  {\bibfnamefont {K.}~\bibnamefont {Watanabe}}, \bibinfo {author}
  {\bibfnamefont {T.}~\bibnamefont {Taniguchi}}, \bibinfo {author}
  {\bibfnamefont {X.}~\bibnamefont {Lu}}, \bibinfo {author} {\bibfnamefont
  {A.~P.}\ \bibnamefont {Higginbotham}}, \emph {et~al.},\ }\bibfield  {title}
  {\bibinfo {title} {Symmetry-broken josephson junctions and superconducting
  diodes in magic-angle twisted bilayer graphene},\ }\href@noop {} {\bibfield
  {journal} {\bibinfo  {journal} {Nature Communications}\ }\textbf {\bibinfo
  {volume} {14}},\ \bibinfo {pages} {2396} (\bibinfo {year}
  {2023})}\BibitemShut {NoStop}%
\bibitem [{\citenamefont {Trahms}\ \emph {et~al.}(2023)\citenamefont {Trahms},
  \citenamefont {Melischek}, \citenamefont {Steiner}, \citenamefont {Mahendru},
  \citenamefont {Tamir}, \citenamefont {Bogdanoff}, \citenamefont {Peters},
  \citenamefont {Reecht}, \citenamefont {Winkelmann}, \citenamefont {von Oppen}
  \emph {et~al.}}]{trahms2023Nature}%
  \BibitemOpen
  \bibfield  {author} {\bibinfo {author} {\bibfnamefont {M.}~\bibnamefont
  {Trahms}}, \bibinfo {author} {\bibfnamefont {L.}~\bibnamefont {Melischek}},
  \bibinfo {author} {\bibfnamefont {J.~F.}\ \bibnamefont {Steiner}}, \bibinfo
  {author} {\bibfnamefont {B.}~\bibnamefont {Mahendru}}, \bibinfo {author}
  {\bibfnamefont {I.}~\bibnamefont {Tamir}}, \bibinfo {author} {\bibfnamefont
  {N.}~\bibnamefont {Bogdanoff}}, \bibinfo {author} {\bibfnamefont
  {O.}~\bibnamefont {Peters}}, \bibinfo {author} {\bibfnamefont
  {G.}~\bibnamefont {Reecht}}, \bibinfo {author} {\bibfnamefont {C.~B.}\
  \bibnamefont {Winkelmann}}, \bibinfo {author} {\bibfnamefont
  {F.}~\bibnamefont {von Oppen}}, \emph {et~al.},\ }\bibfield  {title}
  {\bibinfo {title} {Diode effect in josephson junctions with a single magnetic
  atom},\ }\href@noop {} {\bibfield  {journal} {\bibinfo  {journal} {Nature}\
  }\textbf {\bibinfo {volume} {615}},\ \bibinfo {pages} {628} (\bibinfo {year}
  {2023})}\BibitemShut {NoStop}%
\bibitem [{\citenamefont {Shi}\ \emph {et~al.}(2025)\citenamefont {Shi},
  \citenamefont {Dou}, \citenamefont {Pan}, \citenamefont {Li}, \citenamefont
  {Li}, \citenamefont {Wang}, \citenamefont {Zhang}, \citenamefont {Guo},
  \citenamefont {Deng}, \citenamefont {Tong} \emph {et~al.}}]{JShen2025arXiv}%
  \BibitemOpen
  \bibfield  {author} {\bibinfo {author} {\bibfnamefont {X.}~\bibnamefont
  {Shi}}, \bibinfo {author} {\bibfnamefont {Z.}~\bibnamefont {Dou}}, \bibinfo
  {author} {\bibfnamefont {D.}~\bibnamefont {Pan}}, \bibinfo {author}
  {\bibfnamefont {G.}~\bibnamefont {Li}}, \bibinfo {author} {\bibfnamefont
  {Y.}~\bibnamefont {Li}}, \bibinfo {author} {\bibfnamefont {A.}~\bibnamefont
  {Wang}}, \bibinfo {author} {\bibfnamefont {Z.}~\bibnamefont {Zhang}},
  \bibinfo {author} {\bibfnamefont {X.}~\bibnamefont {Guo}}, \bibinfo {author}
  {\bibfnamefont {X.}~\bibnamefont {Deng}}, \bibinfo {author} {\bibfnamefont
  {B.}~\bibnamefont {Tong}}, \emph {et~al.},\ }\bibfield  {title} {\bibinfo
  {title} {Circuit-level-configurable zero-field superconducting diodes: A
  universal platform beyond intrinsic symmetry breaking},\ }\href@noop {}
  {\bibfield  {journal} {\bibinfo  {journal} {arXiv preprint arXiv:2505.18330}\
  } (\bibinfo {year} {2025})}\BibitemShut {NoStop}%
\bibitem [{\citenamefont {Narita}\ \emph {et~al.}(2022)\citenamefont {Narita},
  \citenamefont {Ishizuka}, \citenamefont {Kawarazaki}, \citenamefont {Kan},
  \citenamefont {Shiota}, \citenamefont {Moriyama}, \citenamefont {Shimakawa},
  \citenamefont {Ognev}, \citenamefont {Samardak}, \citenamefont {Yanase} \emph
  {et~al.}}]{narita2022NN}%
  \BibitemOpen
  \bibfield  {author} {\bibinfo {author} {\bibfnamefont {H.}~\bibnamefont
  {Narita}}, \bibinfo {author} {\bibfnamefont {J.}~\bibnamefont {Ishizuka}},
  \bibinfo {author} {\bibfnamefont {R.}~\bibnamefont {Kawarazaki}}, \bibinfo
  {author} {\bibfnamefont {D.}~\bibnamefont {Kan}}, \bibinfo {author}
  {\bibfnamefont {Y.}~\bibnamefont {Shiota}}, \bibinfo {author} {\bibfnamefont
  {T.}~\bibnamefont {Moriyama}}, \bibinfo {author} {\bibfnamefont
  {Y.}~\bibnamefont {Shimakawa}}, \bibinfo {author} {\bibfnamefont {A.~V.}\
  \bibnamefont {Ognev}}, \bibinfo {author} {\bibfnamefont {A.~S.}\ \bibnamefont
  {Samardak}}, \bibinfo {author} {\bibfnamefont {Y.}~\bibnamefont {Yanase}},
  \emph {et~al.},\ }\bibfield  {title} {\bibinfo {title} {Field-free
  superconducting diode effect in noncentrosymmetric superconductor/ferromagnet
  multilayers},\ }\href@noop {} {\bibfield  {journal} {\bibinfo  {journal}
  {Nature Nanotechnology}\ }\textbf {\bibinfo {volume} {17}},\ \bibinfo {pages}
  {823} (\bibinfo {year} {2022})}\BibitemShut {NoStop}%
\bibitem [{\citenamefont {Lin}\ \emph {et~al.}(2022)\citenamefont {Lin},
  \citenamefont {Siriviboon}, \citenamefont {Scammell}, \citenamefont {Liu},
  \citenamefont {Rhodes}, \citenamefont {Watanabe}, \citenamefont {Taniguchi},
  \citenamefont {Hone}, \citenamefont {Scheurer},\ and\ \citenamefont
  {Li}}]{lin2022NP}%
  \BibitemOpen
  \bibfield  {author} {\bibinfo {author} {\bibfnamefont {J.-X.}\ \bibnamefont
  {Lin}}, \bibinfo {author} {\bibfnamefont {P.}~\bibnamefont {Siriviboon}},
  \bibinfo {author} {\bibfnamefont {H.~D.}\ \bibnamefont {Scammell}}, \bibinfo
  {author} {\bibfnamefont {S.}~\bibnamefont {Liu}}, \bibinfo {author}
  {\bibfnamefont {D.}~\bibnamefont {Rhodes}}, \bibinfo {author} {\bibfnamefont
  {K.}~\bibnamefont {Watanabe}}, \bibinfo {author} {\bibfnamefont
  {T.}~\bibnamefont {Taniguchi}}, \bibinfo {author} {\bibfnamefont
  {J.}~\bibnamefont {Hone}}, \bibinfo {author} {\bibfnamefont {M.~S.}\
  \bibnamefont {Scheurer}},\ and\ \bibinfo {author} {\bibfnamefont
  {J.}~\bibnamefont {Li}},\ }\bibfield  {title} {\bibinfo {title} {Zero-field
  superconducting diode effect in small-twist-angle trilayer graphene},\
  }\href@noop {} {\bibfield  {journal} {\bibinfo  {journal} {Nature Physics}\
  }\textbf {\bibinfo {volume} {18}},\ \bibinfo {pages} {1221} (\bibinfo {year}
  {2022})}\BibitemShut {NoStop}%
\bibitem [{\citenamefont {Wan}\ \emph {et~al.}(2024)\citenamefont {Wan},
  \citenamefont {Qiu}, \citenamefont {Ren}, \citenamefont {Qian}, \citenamefont
  {Li}, \citenamefont {Xu}, \citenamefont {Zhou}, \citenamefont {Zhou},
  \citenamefont {Zhou}, \citenamefont {Wang} \emph {et~al.}}]{wan2024Nature}%
  \BibitemOpen
  \bibfield  {author} {\bibinfo {author} {\bibfnamefont {Z.}~\bibnamefont
  {Wan}}, \bibinfo {author} {\bibfnamefont {G.}~\bibnamefont {Qiu}}, \bibinfo
  {author} {\bibfnamefont {H.}~\bibnamefont {Ren}}, \bibinfo {author}
  {\bibfnamefont {Q.}~\bibnamefont {Qian}}, \bibinfo {author} {\bibfnamefont
  {Y.}~\bibnamefont {Li}}, \bibinfo {author} {\bibfnamefont {D.}~\bibnamefont
  {Xu}}, \bibinfo {author} {\bibfnamefont {J.}~\bibnamefont {Zhou}}, \bibinfo
  {author} {\bibfnamefont {J.}~\bibnamefont {Zhou}}, \bibinfo {author}
  {\bibfnamefont {B.}~\bibnamefont {Zhou}}, \bibinfo {author} {\bibfnamefont
  {L.}~\bibnamefont {Wang}}, \emph {et~al.},\ }\bibfield  {title} {\bibinfo
  {title} {Unconventional superconductivity in chiral molecule--tas2 hybrid
  superlattices},\ }\href@noop {} {\bibfield  {journal} {\bibinfo  {journal}
  {Nature}\ }\textbf {\bibinfo {volume} {632}},\ \bibinfo {pages} {69}
  (\bibinfo {year} {2024})}\BibitemShut {NoStop}%
\bibitem [{\citenamefont {Le}\ \emph {et~al.}(2024)\citenamefont {Le},
  \citenamefont {Pan}, \citenamefont {Xu}, \citenamefont {Liu}, \citenamefont
  {Wang}, \citenamefont {Lou}, \citenamefont {Yang}, \citenamefont {Wang},
  \citenamefont {Yao}, \citenamefont {Wu} \emph {et~al.}}]{le2024nature}%
  \BibitemOpen
  \bibfield  {author} {\bibinfo {author} {\bibfnamefont {T.}~\bibnamefont
  {Le}}, \bibinfo {author} {\bibfnamefont {Z.}~\bibnamefont {Pan}}, \bibinfo
  {author} {\bibfnamefont {Z.}~\bibnamefont {Xu}}, \bibinfo {author}
  {\bibfnamefont {J.}~\bibnamefont {Liu}}, \bibinfo {author} {\bibfnamefont
  {J.}~\bibnamefont {Wang}}, \bibinfo {author} {\bibfnamefont {Z.}~\bibnamefont
  {Lou}}, \bibinfo {author} {\bibfnamefont {X.}~\bibnamefont {Yang}}, \bibinfo
  {author} {\bibfnamefont {Z.}~\bibnamefont {Wang}}, \bibinfo {author}
  {\bibfnamefont {Y.}~\bibnamefont {Yao}}, \bibinfo {author} {\bibfnamefont
  {C.}~\bibnamefont {Wu}}, \emph {et~al.},\ }\bibfield  {title} {\bibinfo
  {title} {Superconducting diode effect and interference patterns in kagome
  csv3sb5},\ }\href@noop {} {\bibfield  {journal} {\bibinfo  {journal}
  {Nature}\ }\textbf {\bibinfo {volume} {630}},\ \bibinfo {pages} {64}
  (\bibinfo {year} {2024})}\BibitemShut {NoStop}%
\bibitem [{\citenamefont {Qi}\ \emph {et~al.}(2025)\citenamefont {Qi},
  \citenamefont {Ge}, \citenamefont {Ji}, \citenamefont {Ai}, \citenamefont
  {Ma}, \citenamefont {Wang}, \citenamefont {Cui}, \citenamefont {Liu},
  \citenamefont {Wang},\ and\ \citenamefont {Wang}}]{Qi_nature2025}%
  \BibitemOpen
  \bibfield  {author} {\bibinfo {author} {\bibfnamefont {S.}~\bibnamefont
  {Qi}}, \bibinfo {author} {\bibfnamefont {J.}~\bibnamefont {Ge}}, \bibinfo
  {author} {\bibfnamefont {C.}~\bibnamefont {Ji}}, \bibinfo {author}
  {\bibfnamefont {Y.}~\bibnamefont {Ai}}, \bibinfo {author} {\bibfnamefont
  {G.}~\bibnamefont {Ma}}, \bibinfo {author} {\bibfnamefont {Z.}~\bibnamefont
  {Wang}}, \bibinfo {author} {\bibfnamefont {Z.}~\bibnamefont {Cui}}, \bibinfo
  {author} {\bibfnamefont {Y.}~\bibnamefont {Liu}}, \bibinfo {author}
  {\bibfnamefont {Z.}~\bibnamefont {Wang}},\ and\ \bibinfo {author}
  {\bibfnamefont {J.}~\bibnamefont {Wang}},\ }\bibfield  {title} {\bibinfo
  {title} {High-temperature field-free superconducting diode effect in
  high-{Tc} cuprates},\ }\href {https://doi.org/10.1038/s41467-025-55880-4}
  {\bibfield  {journal} {\bibinfo  {journal} {Nature Communications}\ }\textbf
  {\bibinfo {volume} {16}},\ \bibinfo {pages} {531} (\bibinfo {year}
  {2025})}\BibitemShut {NoStop}%
\bibitem [{Note1()}]{Note1}%
  \BibitemOpen
  \bibinfo {note} {In the literature, {SDE} (superconducting diode effect)
  typically refers to the phenomenon in bulk superconductors while that in
  Josephson junctions is termed Josephson diode effect (JDE). Here, we use
  supercurrent diode effect (SDE) for both without distinction, given the
  profound similarity under interchange of Cooper pair momentum $q$ and
  Josephson phase $\phi $.}\BibitemShut {Stop}%
\bibitem [{\citenamefont {Nadeem}\ \emph {et~al.}(2023)\citenamefont {Nadeem},
  \citenamefont {Fuhrer},\ and\ \citenamefont {Wang}}]{nadeem2023NRP}%
  \BibitemOpen
  \bibfield  {author} {\bibinfo {author} {\bibfnamefont {M.}~\bibnamefont
  {Nadeem}}, \bibinfo {author} {\bibfnamefont {M.~S.}\ \bibnamefont {Fuhrer}},\
  and\ \bibinfo {author} {\bibfnamefont {X.}~\bibnamefont {Wang}},\ }\bibfield
  {title} {\bibinfo {title} {The superconducting diode effect},\ }\href@noop {}
  {\bibfield  {journal} {\bibinfo  {journal} {Nature Reviews Physics}\ }\textbf
  {\bibinfo {volume} {5}},\ \bibinfo {pages} {558} (\bibinfo {year}
  {2023})}\BibitemShut {NoStop}%
\bibitem [{\citenamefont {Ma}\ \emph {et~al.}(2025{\natexlab{b}})\citenamefont
  {Ma}, \citenamefont {Zhan},\ and\ \citenamefont {Lin}}]{ma2025APR}%
  \BibitemOpen
  \bibfield  {author} {\bibinfo {author} {\bibfnamefont {J.}~\bibnamefont
  {Ma}}, \bibinfo {author} {\bibfnamefont {R.}~\bibnamefont {Zhan}},\ and\
  \bibinfo {author} {\bibfnamefont {X.}~\bibnamefont {Lin}},\ }\bibfield
  {title} {\bibinfo {title} {Superconducting diode effects: Mechanisms,
  materials and applications},\ }\href@noop {} {\bibfield  {journal} {\bibinfo
  {journal} {Advanced Physics Research}\ }\textbf {\bibinfo {volume} {4}},\
  \bibinfo {pages} {2400180} (\bibinfo {year}
  {2025}{\natexlab{b}})}\BibitemShut {NoStop}%
\bibitem [{\citenamefont {Zhang}\ \emph {et~al.}(2022)\citenamefont {Zhang},
  \citenamefont {Gu}, \citenamefont {Li}, \citenamefont {Hu},\ and\
  \citenamefont {Jiang}}]{zhang2022PRX}%
  \BibitemOpen
  \bibfield  {author} {\bibinfo {author} {\bibfnamefont {Y.}~\bibnamefont
  {Zhang}}, \bibinfo {author} {\bibfnamefont {Y.}~\bibnamefont {Gu}}, \bibinfo
  {author} {\bibfnamefont {P.}~\bibnamefont {Li}}, \bibinfo {author}
  {\bibfnamefont {J.}~\bibnamefont {Hu}},\ and\ \bibinfo {author}
  {\bibfnamefont {K.}~\bibnamefont {Jiang}},\ }\bibfield  {title} {\bibinfo
  {title} {General theory of josephson diodes},\ }\href@noop {} {\bibfield
  {journal} {\bibinfo  {journal} {Phys. Rev. X}\ }\textbf {\bibinfo {volume}
  {12}},\ \bibinfo {pages} {041013} (\bibinfo {year} {2022})}\BibitemShut
  {NoStop}%
\bibitem [{\citenamefont {Yuan}\ and\ \citenamefont
  {Fu}(2022)}]{Yuan_pnas2022}%
  \BibitemOpen
  \bibfield  {author} {\bibinfo {author} {\bibfnamefont {N.~F.~Q.}\
  \bibnamefont {Yuan}}\ and\ \bibinfo {author} {\bibfnamefont {L.}~\bibnamefont
  {Fu}},\ }\bibfield  {title} {\bibinfo {title} {Supercurrent diode effect and
  finite-momentum superconductors},\ }\href
  {https://doi.org/10.1073/pnas.2119548119} {\bibfield  {journal} {\bibinfo
  {journal} {Proceedings of the National Academy of Sciences}\ }\textbf
  {\bibinfo {volume} {119}},\ \bibinfo {pages} {e2119548119} (\bibinfo {year}
  {2022})}\BibitemShut {NoStop}%
\bibitem [{\citenamefont {Daido}\ \emph {et~al.}(2022)\citenamefont {Daido},
  \citenamefont {Ikeda},\ and\ \citenamefont {Yanase}}]{Daido_PRL2022}%
  \BibitemOpen
  \bibfield  {author} {\bibinfo {author} {\bibfnamefont {A.}~\bibnamefont
  {Daido}}, \bibinfo {author} {\bibfnamefont {Y.}~\bibnamefont {Ikeda}},\ and\
  \bibinfo {author} {\bibfnamefont {Y.}~\bibnamefont {Yanase}},\ }\bibfield
  {title} {\bibinfo {title} {Intrinsic superconducting diode effect},\ }\href
  {https://doi.org/10.1103/PhysRevLett.128.037001} {\bibfield  {journal}
  {\bibinfo  {journal} {Phys. Rev. Lett.}\ }\textbf {\bibinfo {volume} {128}},\
  \bibinfo {pages} {037001} (\bibinfo {year} {2022})}\BibitemShut {NoStop}%
\bibitem [{\citenamefont {He}\ \emph {et~al.}(2022)\citenamefont {He},
  \citenamefont {Tanaka},\ and\ \citenamefont {Nagaosa}}]{He_NJP2022}%
  \BibitemOpen
  \bibfield  {author} {\bibinfo {author} {\bibfnamefont {J.~J.}\ \bibnamefont
  {He}}, \bibinfo {author} {\bibfnamefont {Y.}~\bibnamefont {Tanaka}},\ and\
  \bibinfo {author} {\bibfnamefont {N.}~\bibnamefont {Nagaosa}},\ }\bibfield
  {title} {\bibinfo {title} {A phenomenological theory of superconductor
  diodes},\ }\href {https://doi.org/10.1088/1367-2630/ac6766} {\bibfield
  {journal} {\bibinfo  {journal} {New Journal of Physics}\ }\textbf {\bibinfo
  {volume} {24}},\ \bibinfo {pages} {053014} (\bibinfo {year}
  {2022})}\BibitemShut {NoStop}%
\bibitem [{\citenamefont {Tanaka}\ \emph {et~al.}(2022)\citenamefont {Tanaka},
  \citenamefont {Lu},\ and\ \citenamefont {Nagaosa}}]{tanaka2022PRB}%
  \BibitemOpen
  \bibfield  {author} {\bibinfo {author} {\bibfnamefont {Y.}~\bibnamefont
  {Tanaka}}, \bibinfo {author} {\bibfnamefont {B.}~\bibnamefont {Lu}},\ and\
  \bibinfo {author} {\bibfnamefont {N.}~\bibnamefont {Nagaosa}},\ }\bibfield
  {title} {\bibinfo {title} {Theory of giant diode effect in d-wave
  superconductor junctions on the surface of a topological insulator},\
  }\href@noop {} {\bibfield  {journal} {\bibinfo  {journal} {Phys. Rev. B}\
  }\textbf {\bibinfo {volume} {106}},\ \bibinfo {pages} {214524} (\bibinfo
  {year} {2022})}\BibitemShut {NoStop}%
\bibitem [{\citenamefont {Lu}\ \emph {et~al.}(2023)\citenamefont {Lu},
  \citenamefont {Ikegaya}, \citenamefont {Burset}, \citenamefont {Tanaka},\
  and\ \citenamefont {Nagaosa}}]{lu2023PRL}%
  \BibitemOpen
  \bibfield  {author} {\bibinfo {author} {\bibfnamefont {B.}~\bibnamefont
  {Lu}}, \bibinfo {author} {\bibfnamefont {S.}~\bibnamefont {Ikegaya}},
  \bibinfo {author} {\bibfnamefont {P.}~\bibnamefont {Burset}}, \bibinfo
  {author} {\bibfnamefont {Y.}~\bibnamefont {Tanaka}},\ and\ \bibinfo {author}
  {\bibfnamefont {N.}~\bibnamefont {Nagaosa}},\ }\bibfield  {title} {\bibinfo
  {title} {Tunable josephson diode effect on the surface of topological
  insulators},\ }\href@noop {} {\bibfield  {journal} {\bibinfo  {journal}
  {Phys. Rev. Lett.}\ }\textbf {\bibinfo {volume} {131}},\ \bibinfo {pages}
  {096001} (\bibinfo {year} {2023})}\BibitemShut {NoStop}%
\bibitem [{\citenamefont {He}\ \emph {et~al.}(2023)\citenamefont {He},
  \citenamefont {Tanaka},\ and\ \citenamefont {Nagaosa}}]{he2023NC}%
  \BibitemOpen
  \bibfield  {author} {\bibinfo {author} {\bibfnamefont {J.~J.}\ \bibnamefont
  {He}}, \bibinfo {author} {\bibfnamefont {Y.}~\bibnamefont {Tanaka}},\ and\
  \bibinfo {author} {\bibfnamefont {N.}~\bibnamefont {Nagaosa}},\ }\bibfield
  {title} {\bibinfo {title} {The supercurrent diode effect and nonreciprocal
  paraconductivity due to the chiral structure of nanotubes},\ }\href@noop {}
  {\bibfield  {journal} {\bibinfo  {journal} {Nature Communications}\ }\textbf
  {\bibinfo {volume} {14}},\ \bibinfo {pages} {3330} (\bibinfo {year}
  {2023})}\BibitemShut {NoStop}%
\bibitem [{\citenamefont {Li}\ and\ \citenamefont {He}(2025)}]{Li_PRB2025}%
  \BibitemOpen
  \bibfield  {author} {\bibinfo {author} {\bibfnamefont {C.}~\bibnamefont
  {Li}}\ and\ \bibinfo {author} {\bibfnamefont {J.~J.}\ \bibnamefont {He}},\
  }\bibfield  {title} {\bibinfo {title} {Microscopic study of supercurrent
  diode effect in chiral nanotubes},\ }\href
  {https://doi.org/10.1103/PhysRevB.111.224504} {\bibfield  {journal} {\bibinfo
   {journal} {Phys. Rev. B}\ }\textbf {\bibinfo {volume} {111}},\ \bibinfo
  {pages} {224504} (\bibinfo {year} {2025})}\BibitemShut {NoStop}%
\bibitem [{\citenamefont {Hou}\ \emph {et~al.}(2023{\natexlab{b}})\citenamefont
  {Hou}, \citenamefont {Nichele}, \citenamefont {Chi}, \citenamefont
  {Lodesani}, \citenamefont {Wu}, \citenamefont {Ritter}, \citenamefont
  {Haxell}, \citenamefont {Davydova}, \citenamefont {Ili\ifmmode~\acute{c}\else
  \'{c}\fi{}}, \citenamefont {Glezakou-Elbert}, \citenamefont {Varambally},
  \citenamefont {Bergeret}, \citenamefont {Kamra}, \citenamefont {Fu},
  \citenamefont {Lee},\ and\ \citenamefont {Moodera}}]{Hou_PRL2023}%
  \BibitemOpen
  \bibfield  {author} {\bibinfo {author} {\bibfnamefont {Y.}~\bibnamefont
  {Hou}}, \bibinfo {author} {\bibfnamefont {F.}~\bibnamefont {Nichele}},
  \bibinfo {author} {\bibfnamefont {H.}~\bibnamefont {Chi}}, \bibinfo {author}
  {\bibfnamefont {A.}~\bibnamefont {Lodesani}}, \bibinfo {author}
  {\bibfnamefont {Y.}~\bibnamefont {Wu}}, \bibinfo {author} {\bibfnamefont
  {M.~F.}\ \bibnamefont {Ritter}}, \bibinfo {author} {\bibfnamefont {D.~Z.}\
  \bibnamefont {Haxell}}, \bibinfo {author} {\bibfnamefont {M.}~\bibnamefont
  {Davydova}}, \bibinfo {author} {\bibfnamefont {S.}~\bibnamefont
  {Ili\ifmmode~\acute{c}\else \'{c}\fi{}}}, \bibinfo {author} {\bibfnamefont
  {O.}~\bibnamefont {Glezakou-Elbert}}, \bibinfo {author} {\bibfnamefont
  {A.}~\bibnamefont {Varambally}}, \bibinfo {author} {\bibfnamefont {F.~S.}\
  \bibnamefont {Bergeret}}, \bibinfo {author} {\bibfnamefont {A.}~\bibnamefont
  {Kamra}}, \bibinfo {author} {\bibfnamefont {L.}~\bibnamefont {Fu}}, \bibinfo
  {author} {\bibfnamefont {P.~A.}\ \bibnamefont {Lee}},\ and\ \bibinfo {author}
  {\bibfnamefont {J.~S.}\ \bibnamefont {Moodera}},\ }\bibfield  {title}
  {\bibinfo {title} {Ubiquitous superconducting diode effect in superconductor
  thin films},\ }\href {https://doi.org/10.1103/PhysRevLett.131.027001}
  {\bibfield  {journal} {\bibinfo  {journal} {Phys. Rev. Lett.}\ }\textbf
  {\bibinfo {volume} {131}},\ \bibinfo {pages} {027001} (\bibinfo {year}
  {2023}{\natexlab{b}})}\BibitemShut {NoStop}%
\bibitem [{\citenamefont {Davydova}\ \emph {et~al.}(2022)\citenamefont
  {Davydova}, \citenamefont {Prembabu},\ and\ \citenamefont
  {Fu}}]{davydova2022SA}%
  \BibitemOpen
  \bibfield  {author} {\bibinfo {author} {\bibfnamefont {M.}~\bibnamefont
  {Davydova}}, \bibinfo {author} {\bibfnamefont {S.}~\bibnamefont {Prembabu}},\
  and\ \bibinfo {author} {\bibfnamefont {L.}~\bibnamefont {Fu}},\ }\bibfield
  {title} {\bibinfo {title} {Universal josephson diode effect},\ }\href@noop {}
  {\bibfield  {journal} {\bibinfo  {journal} {Science advances}\ }\textbf
  {\bibinfo {volume} {8}},\ \bibinfo {pages} {eabo0309} (\bibinfo {year}
  {2022})}\BibitemShut {NoStop}%
\bibitem [{\citenamefont {Xie}\ and\ \citenamefont {Law}(2023)}]{xie2023PRL}%
  \BibitemOpen
  \bibfield  {author} {\bibinfo {author} {\bibfnamefont {Y.-M.}\ \bibnamefont
  {Xie}}\ and\ \bibinfo {author} {\bibfnamefont {K.}~\bibnamefont {Law}},\
  }\bibfield  {title} {\bibinfo {title} {Orbital fulde-ferrell pairing state in
  moir{\'e} ising superconductors},\ }\href@noop {} {\bibfield  {journal}
  {\bibinfo  {journal} {Phys. Rev. Lett.}\ }\textbf {\bibinfo {volume} {131}},\
  \bibinfo {pages} {016001} (\bibinfo {year} {2023})}\BibitemShut {NoStop}%
\bibitem [{\citenamefont {Chakraborty}\ and\ \citenamefont
  {Black-Schaffer}(2025)}]{chakraborty2025prl}%
  \BibitemOpen
  \bibfield  {author} {\bibinfo {author} {\bibfnamefont {D.}~\bibnamefont
  {Chakraborty}}\ and\ \bibinfo {author} {\bibfnamefont {A.~M.}\ \bibnamefont
  {Black-Schaffer}},\ }\bibfield  {title} {\bibinfo {title} {Perfect
  superconducting diode effect in altermagnets},\ }\href@noop {} {\bibfield
  {journal} {\bibinfo  {journal} {Phys. Rev. Lett.}\ }\textbf {\bibinfo
  {volume} {135}},\ \bibinfo {pages} {026001} (\bibinfo {year}
  {2025})}\BibitemShut {NoStop}%
\bibitem [{\citenamefont {Mei}\ \emph {et~al.}(2025)\citenamefont {Mei},
  \citenamefont {Qin},\ and\ \citenamefont {Hu}}]{mei2025interband}%
  \BibitemOpen
  \bibfield  {author} {\bibinfo {author} {\bibfnamefont {J.}~\bibnamefont
  {Mei}}, \bibinfo {author} {\bibfnamefont {S.}~\bibnamefont {Qin}},\ and\
  \bibinfo {author} {\bibfnamefont {J.}~\bibnamefont {Hu}},\ }\bibfield
  {title} {\bibinfo {title} {Interband-pairing-boosted supercurrent diode
  effect in multiband superconductors},\ }\href@noop {} {\bibfield  {journal}
  {\bibinfo  {journal} {arXiv preprint arXiv:2510.15788}\ } (\bibinfo {year}
  {2025})}\BibitemShut {NoStop}%
\bibitem [{\citenamefont {Wang}\ \emph {et~al.}(2012)\citenamefont {Wang},
  \citenamefont {Liang},\ and\ \citenamefont {Hu}}]{WZ_2012arXiv}%
  \BibitemOpen
  \bibfield  {author} {\bibinfo {author} {\bibfnamefont {Z.}~\bibnamefont
  {Wang}}, \bibinfo {author} {\bibfnamefont {Q.-F.}\ \bibnamefont {Liang}},\
  and\ \bibinfo {author} {\bibfnamefont {X.}~\bibnamefont {Hu}},\ }\bibfield
  {title} {\bibinfo {title} {Ratchet potential and rectification effect in
  majorana fermion squid},\ }\href@noop {} {\bibfield  {journal} {\bibinfo
  {journal} {arXiv preprint arXiv:1204.5616}\ } (\bibinfo {year}
  {2012})}\BibitemShut {NoStop}%
\bibitem [{\citenamefont {Chen}\ \emph {et~al.}(2018)\citenamefont {Chen},
  \citenamefont {He}, \citenamefont {Ali}, \citenamefont {Lee}, \citenamefont
  {Fong},\ and\ \citenamefont {Law}}]{CZ_PRB2018}%
  \BibitemOpen
  \bibfield  {author} {\bibinfo {author} {\bibfnamefont {C.-Z.}\ \bibnamefont
  {Chen}}, \bibinfo {author} {\bibfnamefont {J.~J.}\ \bibnamefont {He}},
  \bibinfo {author} {\bibfnamefont {M.~N.}\ \bibnamefont {Ali}}, \bibinfo
  {author} {\bibfnamefont {G.-H.}\ \bibnamefont {Lee}}, \bibinfo {author}
  {\bibfnamefont {K.~C.}\ \bibnamefont {Fong}},\ and\ \bibinfo {author}
  {\bibfnamefont {K.}~\bibnamefont {Law}},\ }\bibfield  {title} {\bibinfo
  {title} {Asymmetric josephson effect in inversion symmetry breaking
  topological materials},\ }\href@noop {} {\bibfield  {journal} {\bibinfo
  {journal} {Phys. Rev. B}\ }\textbf {\bibinfo {volume} {98}},\ \bibinfo
  {pages} {075430} (\bibinfo {year} {2018})}\BibitemShut {NoStop}%
\bibitem [{\citenamefont {Souto}\ \emph {et~al.}(2022)\citenamefont {Souto},
  \citenamefont {Leijnse},\ and\ \citenamefont {Schrade}}]{souto_PRL2022}%
  \BibitemOpen
  \bibfield  {author} {\bibinfo {author} {\bibfnamefont {R.~S.}\ \bibnamefont
  {Souto}}, \bibinfo {author} {\bibfnamefont {M.}~\bibnamefont {Leijnse}},\
  and\ \bibinfo {author} {\bibfnamefont {C.}~\bibnamefont {Schrade}},\
  }\bibfield  {title} {\bibinfo {title} {Josephson diode effect in supercurrent
  interferometers},\ }\href@noop {} {\bibfield  {journal} {\bibinfo  {journal}
  {Phys. Rev. Lett.}\ }\textbf {\bibinfo {volume} {129}},\ \bibinfo {pages}
  {267702} (\bibinfo {year} {2022})}\BibitemShut {NoStop}%
\bibitem [{\citenamefont {Hu}\ \emph {et~al.}(2023)\citenamefont {Hu},
  \citenamefont {Sun}, \citenamefont {Xie},\ and\ \citenamefont
  {Law}}]{hu2023prl}%
  \BibitemOpen
  \bibfield  {author} {\bibinfo {author} {\bibfnamefont {J.-X.}\ \bibnamefont
  {Hu}}, \bibinfo {author} {\bibfnamefont {Z.-T.}\ \bibnamefont {Sun}},
  \bibinfo {author} {\bibfnamefont {Y.-M.}\ \bibnamefont {Xie}},\ and\ \bibinfo
  {author} {\bibfnamefont {K.}~\bibnamefont {Law}},\ }\bibfield  {title}
  {\bibinfo {title} {Josephson diode effect induced by valley polarization in
  twisted bilayer graphene},\ }\href@noop {} {\bibfield  {journal} {\bibinfo
  {journal} {Phys. Rev. Lett.}\ }\textbf {\bibinfo {volume} {130}},\ \bibinfo
  {pages} {266003} (\bibinfo {year} {2023})}\BibitemShut {NoStop}%
\bibitem [{\citenamefont {Shen}\ and\ \citenamefont
  {Zhang}(2025)}]{ShenPRB2025}%
  \BibitemOpen
  \bibfield  {author} {\bibinfo {author} {\bibfnamefont {Q.-K.}\ \bibnamefont
  {Shen}}\ and\ \bibinfo {author} {\bibfnamefont {Y.}~\bibnamefont {Zhang}},\
  }\bibfield  {title} {\bibinfo {title} {Josephson diodes induced by loop
  current states},\ }\href {https://doi.org/10.1103/PhysRevB.111.174515}
  {\bibfield  {journal} {\bibinfo  {journal} {Phys. Rev. B}\ }\textbf {\bibinfo
  {volume} {111}},\ \bibinfo {pages} {174515} (\bibinfo {year}
  {2025})}\BibitemShut {NoStop}%
\bibitem [{\citenamefont {Zinkl}\ \emph {et~al.}(2022)\citenamefont {Zinkl},
  \citenamefont {Hamamoto},\ and\ \citenamefont {Sigrist}}]{zinkl2022prr}%
  \BibitemOpen
  \bibfield  {author} {\bibinfo {author} {\bibfnamefont {B.}~\bibnamefont
  {Zinkl}}, \bibinfo {author} {\bibfnamefont {K.}~\bibnamefont {Hamamoto}},\
  and\ \bibinfo {author} {\bibfnamefont {M.}~\bibnamefont {Sigrist}},\
  }\bibfield  {title} {\bibinfo {title} {Symmetry conditions for the
  superconducting diode effect in chiral superconductors},\ }\href@noop {}
  {\bibfield  {journal} {\bibinfo  {journal} {Phys. Rev. Research}\ }\textbf
  {\bibinfo {volume} {4}},\ \bibinfo {pages} {033167} (\bibinfo {year}
  {2022})}\BibitemShut {NoStop}%
\bibitem [{Note2()}]{Note2}%
  \BibitemOpen
  \bibinfo {note} {We follow the convention of Ref. \cite {buzdin2003prb} to
  call a \protect \text {Josephson} junction with double energy mimina at $\phi
  =\pm \varphi $ a \protect \emph {$\varphi $-junction}, while that with a
  single minimum at $\phi =\varphi _0\protect \neq 0 $ or $ \pi $ a \protect
  \emph {$\varphi _0$-junction}.}\BibitemShut {Stop}%
\bibitem [{\citenamefont {Yang}\ \emph {et~al.}(2018)\citenamefont {Yang},
  \citenamefont {Qin}, \citenamefont {Zhang}, \citenamefont {Fang},\ and\
  \citenamefont {Hu}}]{yang2018PRB}%
  \BibitemOpen
  \bibfield  {author} {\bibinfo {author} {\bibfnamefont {Z.}~\bibnamefont
  {Yang}}, \bibinfo {author} {\bibfnamefont {S.}~\bibnamefont {Qin}}, \bibinfo
  {author} {\bibfnamefont {Q.}~\bibnamefont {Zhang}}, \bibinfo {author}
  {\bibfnamefont {C.}~\bibnamefont {Fang}},\ and\ \bibinfo {author}
  {\bibfnamefont {J.}~\bibnamefont {Hu}},\ }\bibfield  {title} {\bibinfo
  {title} {$\pi$/2-josephson junction as a topological superconductor},\
  }\href@noop {} {\bibfield  {journal} {\bibinfo  {journal} {Phys. Rev. B}\
  }\textbf {\bibinfo {volume} {98}},\ \bibinfo {pages} {104515} (\bibinfo
  {year} {2018})}\BibitemShut {NoStop}%
\bibitem [{\citenamefont {Can}\ \emph {et~al.}(2021)\citenamefont {Can},
  \citenamefont {Tummuru}, \citenamefont {Day}, \citenamefont {Elfimov},
  \citenamefont {Damascelli},\ and\ \citenamefont {Franz}}]{can2021NP}%
  \BibitemOpen
  \bibfield  {author} {\bibinfo {author} {\bibfnamefont {O.}~\bibnamefont
  {Can}}, \bibinfo {author} {\bibfnamefont {T.}~\bibnamefont {Tummuru}},
  \bibinfo {author} {\bibfnamefont {R.~P.}\ \bibnamefont {Day}}, \bibinfo
  {author} {\bibfnamefont {I.}~\bibnamefont {Elfimov}}, \bibinfo {author}
  {\bibfnamefont {A.}~\bibnamefont {Damascelli}},\ and\ \bibinfo {author}
  {\bibfnamefont {M.}~\bibnamefont {Franz}},\ }\bibfield  {title} {\bibinfo
  {title} {High-temperature topological superconductivity in twisted
  double-layer copper oxides},\ }\href@noop {} {\bibfield  {journal} {\bibinfo
  {journal} {Nature Physics}\ }\textbf {\bibinfo {volume} {17}},\ \bibinfo
  {pages} {519} (\bibinfo {year} {2021})}\BibitemShut {NoStop}%
\bibitem [{\citenamefont {Zhao}\ \emph {et~al.}(2023)\citenamefont {Zhao},
  \citenamefont {Cui}, \citenamefont {Volkov}, \citenamefont {Yoo},
  \citenamefont {Lee}, \citenamefont {Gardener}, \citenamefont {Akey},
  \citenamefont {Engelke}, \citenamefont {Ronen}, \citenamefont {Zhong} \emph
  {et~al.}}]{zhao2023Science}%
  \BibitemOpen
  \bibfield  {author} {\bibinfo {author} {\bibfnamefont {S.~F.}\ \bibnamefont
  {Zhao}}, \bibinfo {author} {\bibfnamefont {X.}~\bibnamefont {Cui}}, \bibinfo
  {author} {\bibfnamefont {P.~A.}\ \bibnamefont {Volkov}}, \bibinfo {author}
  {\bibfnamefont {H.}~\bibnamefont {Yoo}}, \bibinfo {author} {\bibfnamefont
  {S.}~\bibnamefont {Lee}}, \bibinfo {author} {\bibfnamefont {J.~A.}\
  \bibnamefont {Gardener}}, \bibinfo {author} {\bibfnamefont {A.~J.}\
  \bibnamefont {Akey}}, \bibinfo {author} {\bibfnamefont {R.}~\bibnamefont
  {Engelke}}, \bibinfo {author} {\bibfnamefont {Y.}~\bibnamefont {Ronen}},
  \bibinfo {author} {\bibfnamefont {R.}~\bibnamefont {Zhong}}, \emph {et~al.},\
  }\bibfield  {title} {\bibinfo {title} {Time-reversal symmetry breaking
  superconductivity between twisted cuprate superconductors},\ }\href@noop {}
  {\bibfield  {journal} {\bibinfo  {journal} {Science}\ }\textbf {\bibinfo
  {volume} {382}},\ \bibinfo {pages} {1422} (\bibinfo {year}
  {2023})}\BibitemShut {NoStop}%
\bibitem [{\citenamefont {Volkov}\ \emph {et~al.}(2024)\citenamefont {Volkov},
  \citenamefont {Lantagne-Hurtubise}, \citenamefont {Tummuru}, \citenamefont
  {Plugge}, \citenamefont {Pixley},\ and\ \citenamefont
  {Franz}}]{volkov2024PRB}%
  \BibitemOpen
  \bibfield  {author} {\bibinfo {author} {\bibfnamefont {P.~A.}\ \bibnamefont
  {Volkov}}, \bibinfo {author} {\bibfnamefont {{\'E}.}~\bibnamefont
  {Lantagne-Hurtubise}}, \bibinfo {author} {\bibfnamefont {T.}~\bibnamefont
  {Tummuru}}, \bibinfo {author} {\bibfnamefont {S.}~\bibnamefont {Plugge}},
  \bibinfo {author} {\bibfnamefont {J.}~\bibnamefont {Pixley}},\ and\ \bibinfo
  {author} {\bibfnamefont {M.}~\bibnamefont {Franz}},\ }\bibfield  {title}
  {\bibinfo {title} {Josephson diode effects in twisted nodal
  superconductors},\ }\href@noop {} {\bibfield  {journal} {\bibinfo  {journal}
  {Phys. Rev. B}\ }\textbf {\bibinfo {volume} {109}},\ \bibinfo {pages}
  {094518} (\bibinfo {year} {2024})}\BibitemShut {NoStop}%
\bibitem [{\citenamefont {Hu}\ \emph {et~al.}(2007)\citenamefont {Hu},
  \citenamefont {Wu},\ and\ \citenamefont {Dai}}]{hu2007prl}%
  \BibitemOpen
  \bibfield  {author} {\bibinfo {author} {\bibfnamefont {J.}~\bibnamefont
  {Hu}}, \bibinfo {author} {\bibfnamefont {C.}~\bibnamefont {Wu}},\ and\
  \bibinfo {author} {\bibfnamefont {X.}~\bibnamefont {Dai}},\ }\bibfield
  {title} {\bibinfo {title} {Proposed design of a josephson diode},\
  }\href@noop {} {\bibfield  {journal} {\bibinfo  {journal} {Phys. Rev. Lett.}\
  }\textbf {\bibinfo {volume} {99}},\ \bibinfo {pages} {067004} (\bibinfo
  {year} {2007})}\BibitemShut {NoStop}%
\bibitem [{\citenamefont {Misaki}\ and\ \citenamefont
  {Nagaosa}(2021)}]{misaki2021prb}%
  \BibitemOpen
  \bibfield  {author} {\bibinfo {author} {\bibfnamefont {K.}~\bibnamefont
  {Misaki}}\ and\ \bibinfo {author} {\bibfnamefont {N.}~\bibnamefont
  {Nagaosa}},\ }\bibfield  {title} {\bibinfo {title} {Theory of the
  nonreciprocal josephson effect},\ }\href@noop {} {\bibfield  {journal}
  {\bibinfo  {journal} {Phys. Rev. B}\ }\textbf {\bibinfo {volume} {103}},\
  \bibinfo {pages} {245302} (\bibinfo {year} {2021})}\BibitemShut {NoStop}%
\bibitem [{\citenamefont {Affleck}\ \emph {et~al.}(2000)\citenamefont
  {Affleck}, \citenamefont {Caux},\ and\ \citenamefont
  {Zagoskin}}]{Affleck2000PRB}%
  \BibitemOpen
  \bibfield  {author} {\bibinfo {author} {\bibfnamefont {I.}~\bibnamefont
  {Affleck}}, \bibinfo {author} {\bibfnamefont {J.-S.}\ \bibnamefont {Caux}},\
  and\ \bibinfo {author} {\bibfnamefont {A.~M.}\ \bibnamefont {Zagoskin}},\
  }\bibfield  {title} {\bibinfo {title} {Andreev scattering and josephson
  current in a one-dimensional electron liquid},\ }\href
  {https://doi.org/10.1103/PhysRevB.62.1433} {\bibfield  {journal} {\bibinfo
  {journal} {Phys. Rev. B}\ }\textbf {\bibinfo {volume} {62}},\ \bibinfo
  {pages} {1433} (\bibinfo {year} {2000})}\BibitemShut {NoStop}%
\bibitem [{\citenamefont {Kusakabe}\ and\ \citenamefont
  {Tanaka}(2002)}]{Tanaka2002PhysicaC}%
  \BibitemOpen
  \bibfield  {author} {\bibinfo {author} {\bibfnamefont {K.}~\bibnamefont
  {Kusakabe}}\ and\ \bibinfo {author} {\bibfnamefont {Y.}~\bibnamefont
  {Tanaka}},\ }\bibfield  {title} {\bibinfo {title} {A theoretical study of a $
  \pi$-junction realized in a quantum dot array},\ }\href
  {https://doi.org/https://doi.org/10.1016/S0921-4534(01)00984-4} {\bibfield
  {journal} {\bibinfo  {journal} {Physica C: Superconductivity}\ }\textbf
  {\bibinfo {volume} {367}},\ \bibinfo {pages} {123} (\bibinfo {year}
  {2002})}\BibitemShut {NoStop}%
\bibitem [{\citenamefont {Bergeret}\ \emph {et~al.}(2007)\citenamefont
  {Bergeret}, \citenamefont {Yeyati},\ and\ \citenamefont
  {Mart\'{\i}n-Rodero}}]{Bergeret2007PRB}%
  \BibitemOpen
  \bibfield  {author} {\bibinfo {author} {\bibfnamefont {F.~S.}\ \bibnamefont
  {Bergeret}}, \bibinfo {author} {\bibfnamefont {A.~L.}\ \bibnamefont
  {Yeyati}},\ and\ \bibinfo {author} {\bibfnamefont {A.}~\bibnamefont
  {Mart\'{\i}n-Rodero}},\ }\bibfield  {title} {\bibinfo {title} {Josephson
  effect through a quantum dot array},\ }\href
  {https://doi.org/10.1103/PhysRevB.76.174510} {\bibfield  {journal} {\bibinfo
  {journal} {Phys. Rev. B}\ }\textbf {\bibinfo {volume} {76}},\ \bibinfo
  {pages} {174510} (\bibinfo {year} {2007})}\BibitemShut {NoStop}%
\bibitem [{Note3()}]{Note3}%
  \BibitemOpen
  \bibinfo {note} {Z. Zhang and J. J. He, to be published.}\BibitemShut {Stop}%
\bibitem [{sup()}]{supp}%
  \BibitemOpen
  \href@noop {} {}\bibinfo {note} {See Supplementary available at
  ...}\BibitemShut {Stop}%
\bibitem [{\citenamefont {Edelstein}(1995)}]{Edelstein1995PRL}%
  \BibitemOpen
  \bibfield  {author} {\bibinfo {author} {\bibfnamefont {V.~M.}\ \bibnamefont
  {Edelstein}},\ }\bibfield  {title} {\bibinfo {title} {Magnetoelectric effect
  in polar superconductors},\ }\href
  {https://doi.org/10.1103/PhysRevLett.75.2004} {\bibfield  {journal} {\bibinfo
   {journal} {Phys. Rev. Lett.}\ }\textbf {\bibinfo {volume} {75}},\ \bibinfo
  {pages} {2004} (\bibinfo {year} {1995})}\BibitemShut {NoStop}%
\bibitem [{\citenamefont {He}\ \emph {et~al.}(2019)\citenamefont {He},
  \citenamefont {Hiroki}, \citenamefont {Hamamoto},\ and\ \citenamefont
  {Nagaosa}}]{he2019CP}%
  \BibitemOpen
  \bibfield  {author} {\bibinfo {author} {\bibfnamefont {J.~J.}\ \bibnamefont
  {He}}, \bibinfo {author} {\bibfnamefont {K.}~\bibnamefont {Hiroki}}, \bibinfo
  {author} {\bibfnamefont {K.}~\bibnamefont {Hamamoto}},\ and\ \bibinfo
  {author} {\bibfnamefont {N.}~\bibnamefont {Nagaosa}},\ }\bibfield  {title}
  {\bibinfo {title} {Spin supercurrent in two-dimensional superconductors with
  rashba spin-orbit interaction},\ }\href@noop {} {\bibfield  {journal}
  {\bibinfo  {journal} {Communications Physics}\ }\textbf {\bibinfo {volume}
  {2}},\ \bibinfo {pages} {128} (\bibinfo {year} {2019})}\BibitemShut {NoStop}%
\bibitem [{\citenamefont {Buzdin}\ and\ \citenamefont
  {Koshelev}(2003)}]{buzdin2003prb}%
  \BibitemOpen
  \bibfield  {author} {\bibinfo {author} {\bibfnamefont {A.}~\bibnamefont
  {Buzdin}}\ and\ \bibinfo {author} {\bibfnamefont {A.}~\bibnamefont
  {Koshelev}},\ }\bibfield  {title} {\bibinfo {title} {Periodic alternating
  0-and $\pi$-junction structures as realization of $\varphi$-josephson
  junctions},\ }\href@noop {} {\bibfield  {journal} {\bibinfo  {journal} {Phys.
  Rev. B}\ }\textbf {\bibinfo {volume} {67}},\ \bibinfo {pages} {220504}
  (\bibinfo {year} {2003})}\BibitemShut {NoStop}%
\end{thebibliography}%
	

\end{document}